\documentclass[preprint,12pt]{elsarticle}

\usepackage{amssymb}
\usepackage{amsmath}
\usepackage{bm}

\begin{document}

\begin{frontmatter}

\title{Phase-field modeling of fracture with physics-informed deep learning}

\author[1]{M. Manav}
\ead{mmanav@ethz.ch}

\author[2]{R. Molinaro}
\ead{roberto.molinaro@sam.math.ethz.ch}

\author[2]{S. Mishra}
\ead{siddhartha.mishra@sam.math.ethz.ch}
\ead[url]{https://camlab.ethz.ch/}

\author[1]{L. De Lorenzis}
\ead{ldelorenzis@ethz.ch}
\ead[url]{https://compmech.ethz.ch/}

\affiliation[1]{organization={Department of Mechanical and Process Engineering,
		ETH Zurich},
           addressline={Tannenstrasse 3}, 
           city={Zurich},
           postcode={8092}, 
           country={Switzerland}}
  
\affiliation[2]{organization={Seminar for Applied Mathematics, Department of Mathematics, ETH Zurich},
       	addressline={Ramistrasse 101}, 
       	city={Zurich},
       	postcode={8092}, 
       	country={Switzerland}}

\begin{abstract}
We explore the potential of the deep Ritz method to learn complex fracture processes such as quasistatic crack nucleation, propagation, kinking, branching, and coalescence within the unified variational framework of phase-field  modeling of brittle fracture. We elucidate the challenges related to the neural-network-based approximation of the energy landscape, and the ability of an optimization approach to reach the correct energy minimum, and we discuss the choices in the construction and training of the neural network which prove to be critical to accurately and efficiently capture all the relevant fracture phenomena. The developed method is applied to several benchmark problems and the results are shown to be in qualitative and quantitative agreement with the finite element solution. The robustness of the approach is tested by using neural networks with different initializations.
\end{abstract}


\end{frontmatter}

\newcommand{\vect}[1]{\bm{#1}}
\newcommand{\fref}[1]{Figure~\ref{#1}}
\newcommand{\sref}[1]{Section~\ref{#1}}
\newcommand{\tref}[1]{Table~\ref{#1}}

\section{Introduction}
\label{intro}
Predictive modeling and simulation of fracture processes is critically important in design for many engineering applications and still poses significant challenges. In the past two decades, phase-field fracture modeling~\cite{bourdin2000numerical,bourdin2008variational} has emerged as a game changer; it offers a unified variational framework and, upon discretization, a flexible computational method to encompass various aspects of fracture including crack nucleation, propagation, merging, and branching~\cite{bourdin2008variational,ambati2015review,wu2020phase}, while obviating the need for crack tracking procedures as well as for ad-hoc parameters and criteria. For these reasons, it has found applications and extensions to a variety of fracture problems including thermal cracks~\cite{bourdin2014morphogenesis}, ductile fracture~\cite{alessi2018comparison}, drying-induced cracks~\cite{luo2023phase}, and hydraulic fracture~\cite{heider2021review}, among many others. Furthermore, it has been demonstrated to accurately capture topologically complex crack paths, even in three dimensions~\cite{bourdin2014morphogenesis}, while delivering quantitatively accurate predictions~\cite{wu_2017}.

The first variational phase-field model for brittle fracture was derived by regularizing the free-discontinuity problem obtained from recasting Griffith's fracture criterion in a variational form~\cite{francfort1998revisiting, bourdin2000numerical}. Regularization is performed by introducing an additional field, known as the phase field, ranging between 0 (representing the intact material) and 1 (representing the fully cracked material). Localization of this field in a narrow band, whose width is controlled by a length-scale parameter inherent to the model, gives rise to a smeared representation of a crack. Later, it was shown that a more flexible but substantially analogous variational modeling framework can be constructed from a special family of gradient damage models~\cite{pham2011gradient}, whereby the phase field is interpreted as a damage variable and is therefore postulated to be irreversible. In the computational setting, the regularization of a sharp crack to a smeared representation overcomes the computational challenges encountered in handling an a priori unknown crack path. The crack is now represented implicitly by the continuous phase field; along with the displacement field, this field is obtained by solving the non-linear coupled partial differential equations (PDEs) stemming from the stationarity conditions of the governing energy functional. However, two challenging aspects remain: the energy functional is non-convex, which needs special attention in the numerics ~\cite{lorenzis2020numerical}, and the resolution of the (small) length scale in the discretized setting is computationally expensive. This limitation restricts the applicability of the approach to e.g. optimization, uncertainty quantification and inverse problems.

In recent years, considerable effort has been directed towards leveraging advances in machine learning to accurately model physical phenomena~\cite{karniadakis2021physics}; various physics-informed deep learning approaches~\cite{lagaris1998artificial} have been employed to solve PDEs and variational problems~\cite{sirignano2018dgm,raissi2019physics,kharazmi2021hp,yu2018deep}. One prominent approach, known as physics-informed neural networks (PINNs), learns the solution of a PDE by the unsupervised (or semi-supervised) training of a neural network (NN)~\cite{lagaris1998artificial,raissi2019physics,mishra2019,mishra2022estimates}. PINNs build upon the universal approximation property of NNs~\cite{cybenko1989approximation,hornik1989multilayer,yarotsky2017error}, i.e. the ability of NNs to approximate any continuous function, which allows for the use of NNs as the ansatz space for PDE solutions. Notably, PINNs can seamlessly incorporate any available data and compute solutions to forward and inverse problems with the same NN architecture. They have demonstrated accurate approximation of solutions to a variety of PDEs and inverse problems, also in solid mechanics~\cite{haghighat2021physics,rao2021physics,zhang2022analyses,rezaei2022mixed,henkes2022physics,bastek2023physics,niu2023modeling,faroughi2024physics}. Rigorous estimates of the generalization error of PINNs approximations have also been established~\cite{shin2020convergence,mishra2022estimates,mishra2022estimatesInv}. Furthermore, multiple variations of PINNs have been proposed, such as, hp-VPINNs~\cite{kharazmi2021hp}, cPINNs~\cite{jagtap2020conservative}, xPINNS~\cite{jagtap2021extended}, FBPINNs~\cite{moseley2023finite}, among others, to tailor them for different problems and to improve their performance.

In PINNs, training of the NN representing the solution field involves the minimization of a loss function comprising the PDE residual and additional terms, which enforce initial and boundary conditions and possibly other constraints. An alternative approach known as the deep Ritz method (DRM)~\cite{yu2018deep,nguyen2020deep,samaniego2020energy}, directly minimizes the energy functional instead of the PDE residual. The DRM is particularly useful in obtaining stable solutions when the governing energy functional is non-convex, as in phase-field fracture modeling. Additionally, since the computation of the energy requires a lower order of derivatives compared to the corresponding strong-form PDE, the DRM is expected to be computationally less expensive~\cite{nguyen2020deep}. On the other hand, energy calculation requires integration over the entire domain, and only a few studies have investigated quadrature rules for NNs~\cite{berrone2022variational}.

PINN approaches have been developed to solve phase-field problems (Cahn-Hilliard and Allen-Cahn equations) involving the phase field as the only scalar unknown~\cite{wight2020solving,mattey2022novel}. In contrast, phase-field fracture modeling involves two coupled fields: the displacement (vector) field and the phase field. Furthermore, while the steep variation of the phase field at localization is governed by the regularization length scale, the displacement field undergoes a much sharper variation wherever the phase field localizes. These differences lead to additional and unique challenges in approximating the solution of phase-field fracture problems using NNs.  While the DRM has shown some promise, its applicability has been demonstrated primarily in problems involving crack propagation with simple crack paths~\cite{goswami2020transfer,goswami2020adaptive}. Crack initiation in the absence of a notch has only been demonstrated by assuming a very large value for the regularization length scale~\cite{goswami2020transfer}. Additionally, these approaches have required a large number of quadrature points, exceeding the number needed to resolve the regularization length scale. Notably, a quantitative assessment of the solution accuracy is also lacking. The applicability of approaches other than the DRM has been partially investigated in~\cite{goswami2020transfer,ghaffari2022deep}. An operator learning approach~\cite{lu2021learning,li2021physics}, namely variational DeepONets, has also been applied to predict the crack path in a problem involving crack propagation in quasi-brittle materials~\cite{goswami2022physics}. However, the applicability of physics-informed deep learning to phase-field fracture modeling hinges on the ability of NNs to learn all the fracture processes accurately, especially considering that this flexibility lies at the core of the phase-field approach and is also one of the key reasons of its success.

In this work, we elucidate the challenges involved in learning the solution of phase-field fracture problems, and investigate the design of NNs and of their training strategies aimed at learning crack initiation, propagation, kinking, branching, and coalescence. We also address the challenges associated with learning the solution fields with the same level of domain discretization as required in finite element analysis (FEA). We systematically demonstrate the accuracy of the learned solution by comparing it with the FEA solution. Additionally, we test the robustness of our approach by training NNs with different random initializations.

The paper is organized as follows. A brief review of the phase-field brittle fracture formulation is presented in~\sref{PFF}. In~\sref{DL}, the DRM is summarized, and the challenges in learning the solution of a phase-field fracture problem are discussed along with the design of the NNs and their training strategy. This is followed by numerical examples of crack nucleation, propagation, kinking, branching, and coalescence in~\sref{numE}. Finally, we summarize the main conclusions in~\sref{conclusions}.

\section{Phase-field model of brittle fracture}
\label{PFF}
In this section, we summarize the governing equations of the phase-field model of brittle fracture which we aim at solving with the deep learning approach proposed in the following sections. 
\subsection{Basics of the formulation}
Let $\Omega\subset\mathbb{R}^d,~d\in\{1,2,3\}$ be a bounded open domain occupied by a $d$-dimensional body. $\Gamma_{D,0}$, $\Gamma_{D,1}$ and $\Gamma_{N}$ are disjoint sections of the boundary $\partial\Omega$ of $\Omega$ with prescribed homogeneous Dirichlet, non-homogeneous Dirichlet and Neumann boundary conditions, respectively. The body is assumed to be linear elastic with the strain energy density given by $\Psi(\bm{\varepsilon})=\frac{1}{2}\lambda{\rm tr}^2(\bm{\varepsilon})+\mu{\rm tr}(\bm{\varepsilon}\cdot\bm{\varepsilon})$, where $\lambda$ and $\mu$ are the Lam\'e constants, and $\bm{\varepsilon} = \frac{1}{2}(\bm{\nabla u}+\bm{\nabla u}^T)$ is the infinitesimal strain tensor, with $\bm{u}:\Omega\to\mathbb{R}^d$ as the displacement field. The total energy functional for the body acted upon by body force $\bar{\vect{b}}$ and surface traction $\bar{\vect{t}}$ is given by:
\begin{align}
		\mathcal{E}(\vect{u},\alpha)=&\underbrace{\int_{\Omega} \left(g(\alpha)\Psi^+(\bm{\varepsilon}(\vect{u}))+\Psi^-(\bm{\varepsilon}(\vect{u}))\right) {\rm d}\Omega}_{\mathcal{E}^{el}}+
		\underbrace{\frac{G_c}{c_w}\int_{\Omega}\left( \frac{w(\alpha)}{l}+l|\bm{\nabla}\alpha|^2\right) {\rm d}\Omega}_{\mathcal{E}^{d}} \nonumber\\ 
		&-\int_{\Omega}\bar{\vect{b}}\cdot\vect{u}\, {\rm d}\Omega - \int_{\Gamma_{N}}\bar{\vect{t}}\cdot\vect{u}\, {\rm d}s. \label{eq1}
\end{align}
Here $\mathcal{E}^{el}$ and $\mathcal{E}^{d}$ represent the elastic and damage parts of the energy, respectively. $\alpha:\Omega\to[0,~1]$ is the phase field (or damage field), with $\alpha=0$ representing the undamaged state and $\alpha=1$ the fully damaged state. The function $g(\alpha)$, known as degradation function, modulates the degradation of the elastic energy with increasing damage. $w(\alpha)$ is also known as local dissipation function, as it dictates the energy dissipated through damage in the unit volume of the material. The parameter $l$ with $0<l\ll{\rm diam}(\Omega)$ is the regularization length controlling the thickness of the transition zone between undamaged and fully damaged material regions; $G_c$ represents the fracture toughness of the material, and $c_w$ is a normalization constant given by $c_w = 4\int_0^1\sqrt{w(t)}dt$. 
In this work, we adopt the following choices~\cite{pham2011gradient}:
\begin{align}
	&{\rm AT1 \,\,model}: \quad g(\alpha) := (1-\alpha)^2 + \eta,\quad w(\alpha) := \alpha, \quad c_w=8/3, \label{eq2b}\\
	&{\rm AT2\,\, model}: \quad g(\alpha) := (1-\alpha)^2  + \eta,\quad w(\alpha) := \alpha^2, \quad c_w=2. \label{eq2c}
\end{align}
with $\eta=o(l)$.
The elastic strain energy density $\Psi$ is decomposed into active (i.e. damage-driving) and inactive (i.e. damage-resisting) parts, $\Psi^+$ and $\Psi^-$, respectively, and the degradation is applied only to $\Psi^+$~\cite{amor2009regularized,freddi2010regularized,miehe2010thermodynamically,miehe2010phase}. While several energy decompositions are available
~\cite{amor2009regularized,freddi2010regularized,miehe2010thermodynamically,vicentini2024}, in this work we focus on the volumetric-deviatoric decomposition~\cite{amor2009regularized} whereby
\begin{align}
	&\Psi^+=\frac{1}{2}K\langle\rm tr(\bm{\varepsilon})\rangle_+^2+\mu{\rm tr}(\mathbf{e}\cdot\mathbf{e}), \nonumber \\
	& \Psi^-=\frac{1}{2}K\langle\rm tr(\bm{\varepsilon})\rangle_-^2,
	\label{eq2d}
\end{align}
with $\langle a \rangle_+ = {\rm max}\{0, a\}$, $\langle a \rangle_- = {\rm min}\{0, a\}$, the bulk modulus $K=\lambda + \frac{2}{3}\mu$, and the deviatoric strain tensor $\mathbf{e}=\bm{\varepsilon}-\frac{tr(\bm{\varepsilon})}{3}\mathbf{I}$, where $\mathbf{I}$ is the third-order identity tensor.

\subsection{Time-discrete evolution problem}
In the time-discrete setting of the quasi-static evolution problem, the state of the system at the (pseudo-)time or loading step $n\ge1$ is obtained as the solution of the minimization problem
\begin{equation}
	\operatorname*{argmin}_{\vect{u},\alpha} \{ \mathcal{E}_n(\vect{u},\alpha) :~ \vect{u}\in \vect{V}_{\bar{\vect{u}}_n}, ~\alpha\in\mathcal{D}_{\alpha_{n-1}}\},
	\label{eq3}
\end{equation}
where 
\begin{equation}
		\mathcal{E}_n(\vect{u},\alpha)=\mathcal{E}^{el}(\vect{u},\alpha)+
		\mathcal{E}^{d}(\alpha)-\int_{\Omega}\bar{\vect{b}}_n\cdot\vect{u}\, {\rm d}\Omega - \int_{\Gamma_{N}}\bar{\vect{t}}_n\cdot\vect{u}\, {\rm d}s,
\end{equation}
\begin{equation}
	\vect{V}_{\bar{\vect{u}}_n}:=\{\vect{u}\in(H^1(\Omega))^d: ~\vect{u}=\bm{0} ~{\rm on} ~\Gamma_{D,0}, ~\vect{u}=\bar{\vect{u}}_n ~{\rm on} ~\Gamma_{D,1} \}
	\label{eq4a}
\end{equation}
denotes the space of kinematically admissible displacement fields, and
\begin{equation}
	\mathcal{D}_{\alpha_{n-1}}:=\{ \alpha\in H^1(\Omega): ~\alpha\ge \alpha_{n-1} ~{\rm in}~ \Omega \}
	\label{eq4b}
\end{equation}
denotes the space of admissible phase fields. The condition $\alpha\ge \alpha_{n-1}$ represents the irreversibility of the phase field evolution. The first-order necessary conditions of the minimization problem deliver the strong form of the governing equations
\begin{align}
 &\bm{\nabla}\cdot\bm{\sigma}+\bar{\vect{b}}_n=0, \label{eq5a} \\
 \quad &\alpha-\alpha_{n-1} \ge 0, \label{eq5b} \\
&g'(\alpha)\Psi^+ +\frac{G_c}{c_w}\left(\frac{w'(\alpha)}{l}-2 l\Delta\alpha \right)\ge 0, \label{eq5c} \\
& (\alpha-\alpha_{n-1})\left[g'(\alpha)\Psi^+ +\frac{G_c}{c_w}\left(\frac{w'(\alpha)}{l}-2 l\Delta\alpha \right) \right] =0, \label{eq5d}
\end{align}
all valid in $\Omega$, and represent the local equilibrium of forces, damage irreversibility, damage criterion and loading-unloading conditions, respectively. The Cauchy stress is defined as $\bm{\sigma}=g(\alpha)\frac{\partial\Psi^+}{\partial\bm{\varepsilon}}+\frac{\partial\Psi^-}{\partial\bm{\varepsilon}}$, and $()' = \frac{\partial}{\partial\alpha}()$. The whole set of boundary conditions reads as follows:
\begin{align}
	\vect{u} = \vect{0} \quad &{\rm on~} \Gamma_{D,0}, \nonumber \\
	\vect{u} = \bar{\vect{u}}_n \quad &{\rm on~} \Gamma_{D,1}, \nonumber \\
	\bm{\sigma}\vect{n} = \bar{\vect{t}}_n \quad &{\rm on~} \Gamma_{N}, \nonumber \\
	\alpha-\alpha_{n-1} \ge 0,\,\,\nabla\alpha\cdot\vect{n} \ge 0, \,\, (\alpha-\alpha_{n-1})(\nabla\alpha\cdot\vect{n})=0 \quad &{\rm on~} \Gamma.
	\label{eq6}
\end{align}

\subsection{Irreversibility of the phase field}
\label{IRA}
One way to enforce the irreversibility of the phase field in the computations is to add to the energy functional in \eqref{eq1} the following energetic penalty~\cite{gerasimov2019penalization}:
\begin{equation}
	\mathcal{E}^{ir}(\alpha)=\int_{\Omega} \frac{1}{2}\gamma_{ir}\langle \alpha-\alpha_{n-1}\rangle_-^2{\rm d}\Omega, \label{eq6a}
\end{equation}
where $\gamma_{ir}$ is the penalty parameter, whose value can be chosen as suggested in~\cite{gerasimov2019penalization}:
\begin{align}
	&\gamma_{ir}=\frac{G_c}{l}\frac{27}{64\mathtt{TOL_{ir}^2}} \quad\quad\quad\,\, ({\rm AT1~model}), \label{eq2e}\\
	&\gamma_{ir}=\frac{G_c}{l}\left(\frac{1}{\mathtt{TOL_{ir}^2}}-1\right) \quad ({\rm AT2~model}). \label{eq2f}
\end{align}
Here $0<\mathtt{TOL_{ir}}\le 1$ is the prescribed irreversibility tolerance threshold; in this work, we set it to $\mathtt{TOL_{ir}}=5\times10^{-3}$.

\subsection{Non-dimensionalization scheme}
\label{NDS}
Non-dimensionalization not only reduces the number of relevant physical parameters, but also aids in learning by ensuring that the inputs are  $\sim 1$. So, we non-dimensionalize the energy functional using the following scheme:
\begin{align}
	&\tilde{\vect{x}}=\frac{\vect{x}}{L}, \quad \tilde{l}=\frac{l}{L}, \quad \tilde{\vect{u}}=\frac{\vect{u}}{L}\left(\frac{G_c}{E l}\right)^{-1/2}, \nonumber \\ &\tilde{\lambda}=\frac{\lambda}{E}, \quad \tilde{\mu}=\frac{\mu}{E}, \quad {\rm and} \quad \tilde{\mathcal{E}}=\frac{l \mathcal{E}}{L^3G_c},
	\label{eq6c}
\end{align}
where $\vect{x}\in\Omega$ ($\tilde{\vect{x}}\in\tilde{\Omega}$) is a material point coordinate, $L$ is the characteristic length of the body, $E$ is the Young's modulus, and $\sqrt{\frac{G_c}{E l}}$ is of the order of the critical strain for crack nucleation in a 1D bar~\cite{pham2011gradient}. Then the dimensionless form of the energy functional in \eqref{eq1}, with no surface tractions and body forces, and including the energetic penalty to enforce the irreversibility of the phase field, reads
\begin{align}
	\tilde{\mathcal{E}}(\tilde{\vect{u}},\alpha)=&\underbrace{\int_{\tilde{\Omega}} \left(g(\alpha)\tilde{\Psi}^+(\tilde{\bm{\varepsilon}}(\tilde{\vect{u}}))+\tilde{\Psi}^-(\tilde{\bm{\varepsilon}}(\tilde{\vect{u}}))\right) {\rm d}\tilde{\Omega}}_{\tilde{\mathcal{E}}^{el}} \nonumber \\
	&+\underbrace{\frac{1}{c_w}\int_{\tilde{\Omega}}\left( w(\alpha)+\tilde{l}^2|\bm{\tilde{\nabla}}\alpha|^2\right) {\rm d}\tilde{\Omega}}_{\tilde{\mathcal{E}}^{d}} \nonumber \\
    &+\underbrace{\int_{\Omega} \frac{1}{2}\tilde{\gamma}_{ir}\langle \alpha-\alpha_{n-1}\rangle_-^2{\rm d}\tilde{\Omega}}_{\tilde{\mathcal{E}}^{ir}},
	\label{eq6d}
\end{align}
where $\bm{\tilde{\nabla}} = L\bm{\nabla}$, $\tilde{\bm{\varepsilon}}=\bm{\varepsilon}\left(\frac{G_c}{E l}\right)^{-1/2}$, $\tilde{\Psi}^{\pm}=\frac{l{\Psi}^{\pm}}{G_c}$, $\tilde{\gamma}_{ir}=\frac{l\gamma_{ir}}{G_c}$, $\tilde{\mathcal{E}}^{el}=\frac{l \mathcal{E}^{el}}{L^3G_c}$, $\tilde{\mathcal{E}}^{d}=\frac{l \mathcal{E}^{d}}{L^3G_c}$, and $\tilde{\mathcal{E}}^{ir}=\frac{l \mathcal{E}^{ir}}{L^3G_c}$. Also, the non-dimensional stress follows as $\tilde{\bm{\sigma}}=g(\alpha)\frac{\partial \tilde{\Psi}^+}{\partial\tilde{\varepsilon}}+\frac{\partial \tilde{\Psi}^-}{\partial\tilde{\varepsilon}}$. Furthermore, leveraging the relation among Lam\'e constants, Young's modulus and Poisson's ratio $\nu$, we obtain $\tilde{\lambda}=\frac{\nu}{(1+\nu)(1-2\nu)}$ and $\tilde{\mu}=\frac{1}{2(1+\nu)}$. Notably, as a consequence, the only parameters in the non-dimensional energy are $\nu$ and $\tilde{l}$.  For simplicity of notation, from here onwards, we only use non-dimensional quantities but omit the  $\tilde{(\cdot)}$ symbol.

\section{A Deep Ritz method for phase-field fracture modeling}
\label{DL}
\subsection{The Deep Ritz method}
In the DRM, as in any other physics-informed deep learning approach for solving PDEs and variational problems, the solution field is represented by an NN, frequently a feedforward NN (also known as a multilayer perceptron). A feedforward NN is a composition of affine transformations and scalar non-linear activation functions. Given an input $\vect{y}\in\mathbb{R}^n$, the NN maps it to the output $\vect{z}_{\theta}\in \mathbb{R}^m$ as follows
\begin{equation}
	\vect{z}_{\theta}(\vect{y})=C_K\circ\sigma\circ C_{K-1}\cdot\cdot\cdot \circ\,\,\sigma\circ C_1(\vect{y}), \label{eq7}
\end{equation}
where $C_k$ (with $1\le k\le K$) represents an affine transformation, also known as the $k^{th}$ layer of the NN. Specifically, $C_k \vect{z}_k=W_k\vect{z}_k+b_k$, where $W_k$ and $b_k$ are trainable weights and biases and $\vect{z}_k$ is the input to the layer. The symbol $\circ$ represents the composition of functions, and $\sigma$ is a scalar activation function. Commonly used activation functions are the ReLU, sigmoid, and hyperbolic tangent ($\tanh$) functions~\cite{dubey2022activation}. $\theta$ denotes the set of all trainable parameters of the NN, i.e. $\theta=\{W_k,~b_k\}, \forall~1\le k\le K$.

Training of the NN aims to find $\theta$ such that $\vect{z}_{\theta}$ approximates the target solution. It involves constructing an appropriate loss function and then minimizing the loss with respect to the parameters in $\theta$. The minimization is typically performed using stochastic gradient decent algorithms such as Adam~\cite{kingma2014adam} or higher-order optimization algorithms such as L-BFGS~\cite{liu1989limited}.

For our problem, we define the displacement and phase fields obtained from the NN as $\vect{u}_{\theta}(\vect{x})$ and $\alpha_{\theta}(\vect{x})$. Then the loss function is expressed as
\begin{equation}
	\mathcal{L} = \log(\mathcal{E}_{\theta}), \label{eq8a}
\end{equation}
where $\mathcal{E}_{\theta} = \mathcal{E}(\vect{u}_{\theta}, \alpha_{\theta})$ is obtained from \eqref{eq6d}. Taking the $\log$ of the energy in \eqref{eq8a} makes the loss function $\sim 1$, allowing us to set the same weight for weight regularization (to be discussed later) and the same stopping criterion for the optimizer for all the problems investigated in this work. 

Note that, in our experience, the enforcement of the Dirichlet boundary conditions using a soft constraint in DRM is challenging. Hence, we enforce them by constructing an ansatz, following the approach employed in~\cite{lagaris1998artificial}, see Section \ref{arch}. For a general domain shape for which an ansatz is not feasible, general approaches for the enforcement of boundary conditions as in~\cite{berg2018unified,leake2020deep} can be employed.

\subsection{Construction and training of the NN}
\label{NNmodel}
Our aim is to obtain an NN-based approximation of the solution fields employing the DRM in learning. This approach entails two key challenges. The \textit{first challenge} lies in ensuring that the NN-based approximation of the energy surface, $\mathcal{E}_{\theta}$, is a faithful projection of the energy landscape in the space of NN-parameters. Specifically, it is crucial that the energy barriers and minima are appropriately represented. 
In turn, the representation of the energy surface relies on three ingredients: the construction of the NN(s) representing the fields, the computation of the gradients of the fields, and the integration over the domain. 
The \textit{second challenge} lies in the choice of the optimization algorithm. This algorithm needs to respect the energy barriers and reach the correct local energy minimum regardless of the complexity of the solution associated with that energy minimum. As follows, we discuss all these aspects and refer to some Appendices for additional details.

\begin{figure}[!h]
	\begin{center}
		\includegraphics[width=0.65\textwidth]{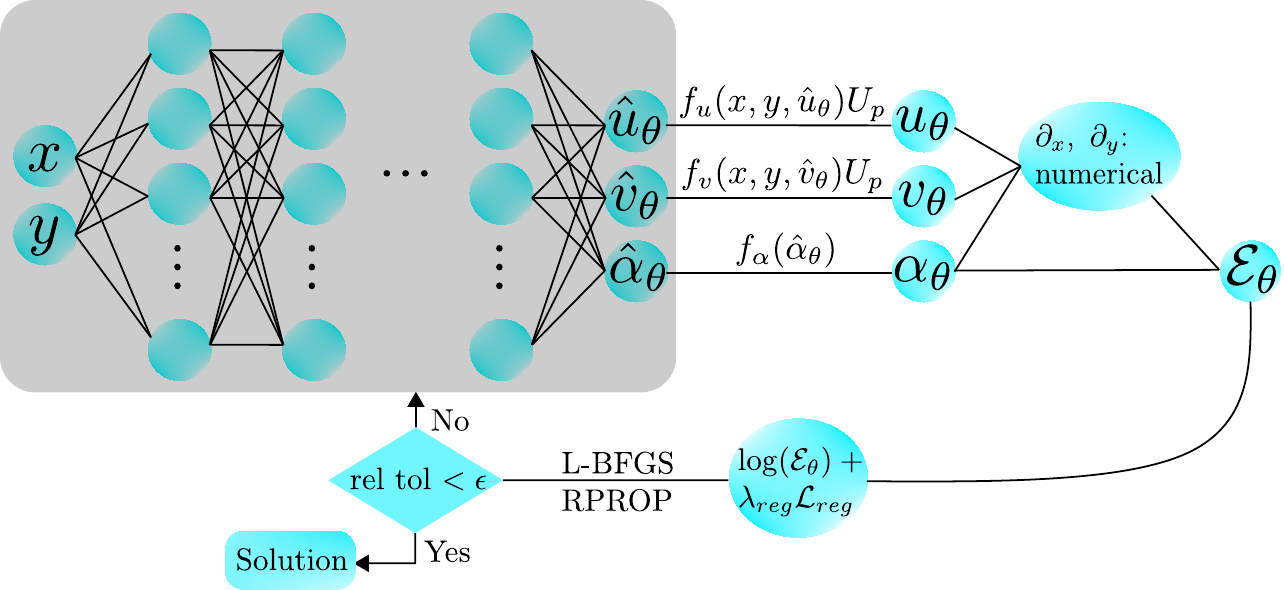}
        \hfill
        \includegraphics[width=0.34\textwidth]{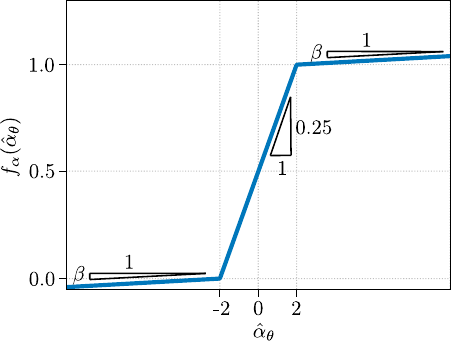}
	\end{center}
	\caption{Scheme of our deep Ritz method for the two-dimensional case (left); function constraining the phase field between 0 and 1 (right)}
	\label{DRZ}
\end{figure}
\subsubsection{Choice of the NN architecture}
\label{arch}
Physically distinct fields such as the displacement field and the phase field are expected to be learned efficiently when represented by independent NNs. However, we find that  an intimate coupling between the fields, facilitated by representing the displacement and phase fields as the outputs of the same NN, helps in learning the nucleation of a crack as shown in~\ref{2NN}. Hence, we construct an NN which takes as input the position vector of a point in the domain (2 components in 2D, 1 in 1D) and delivers outputs which correspond to the components of the displacement vector and the value of the phase field at that point (3 outputs in 2D, 2 in 1D), see~\fref{DRZ}. The non-dimensionalization scheme outlined in~\sref{NDS} ensures that the inputs are $\sim 1$. In constructing an ansatz to apply Dirichlet boundary conditions, we ensure that the outputs of the NN corresponding to the displacement field are also approximately $\sim 1$ as follows:
\begin{align}
    &u_{\theta}=f_u(\vect{x},\hat{u}_{\theta})U_p,\\
    &v_{\theta}=f_v(\vect{x},\hat{v}_{\theta})U_p,
\end{align}
where $\hat{u}_{\theta}$ and $\hat{v}_{\theta}$ are the direct NN outputs corresponding to the displacement components, and $U_p$ is the prescribed displacement. $f_u$ and $f_v$ are designed for each problem. To ensure that $\alpha_{\theta}\in [0,1]$, the application of the sigmoid function to the direct NN output corresponding to the phase field, $\hat{\alpha}_{\theta}$, is an attractive choice; however, it turns out that its vanishing slope hinders crack nucleation. Therefore, we design the following function, plotted in~\fref{DRZ}, to overcome this problem:
\begin{equation}
	\alpha_{\theta}=f_{\alpha}(\hat{\alpha}_{\theta})=\begin{cases}
		\frac{\hat{\alpha}_{\theta}}{4}+\frac{1}{2}, & {\rm for~} |\hat{\alpha}_{\theta}|\le 2\\
		\beta(\hat{\alpha}_{\theta}+2), & {\rm for~} \hat{\alpha}_{\theta}<-2\\
		\beta(\hat{\alpha}_{\theta}-2)+1, & {\rm for~} \hat{\alpha}_{\theta}>2
	\end{cases}
\end{equation}
By setting $\alpha_{\theta}=f_{\alpha}(\hat{\alpha}_{\theta})$ in the 1D bar problem, we find that the order of the smallest $\beta$ for which an NN can learn the correct critical load for crack nucleation is $10^{-3}$. So, we set $\beta=10^{-3}$ in all examples of this paper.

\subsubsection{Choice of the activation function}
Another important aspect of the NN design is the choice of the non-linear activation function. In PINNs, smooth activations like $\tanh$ are frequently employed. However, we find that an NN with this activation fails to learn the correct crack path (see \ref{act}). The evolution of the loss during training suggests that an NN with $\tanh$ activation fails to represent the energy barrier in the loss landscape which would prevent crack propagation in an incorrect direction. Therefore, we employ ReLU activation with the following modification:
\begin{equation}
	z_{k+1}=\max\{0,~m_k (W_k z_k + b_k)\},
\end{equation}
where $m_k$ are learnable coefficients. While initializing the NN, we set $m_k=m_0$, where $m_0$ is a constant. We observe that while the value of $m_k$ does not change significantly during training, some choices of $m_0$, as detailed in~\sref{numSetup}, facilitate learning the critical applied displacement for crack initiation or propagation with higher accuracy.

\subsubsection{Gradient computation and quadrature}
\label{gradquad}
Computation of the gradients poses another challenge. When employing autodifferentiation to compute the gradients of the fields represented by an NN with a smooth activation function like $\tanh$, the learned fields display high-frequency oscillations (similar to those resulting from Gibbs phenomena) leading to the localization of the stress in random directions near the crack tip (see \ref{diff}). This can be viewed as an undesired modification of the energy barrier, allowing for crack propagation in unexpected directions and leading to an inaccurate crack path prediction. 

While this is an additional argument against the use of tanh activation, the problem is not solved by the use of the ReLU activation function. To address it,
 we discretize the domain as we would do in FEA. The field values at the nodes are obtained from the NN, whereas the field values and their gradients within each element, in particular at the Gauss points, are computed using shape functions like in FEA, as follows
\begin{align}
	&{\bm u}^h_{\theta}({\bm x})=\sum_{a=1}^n N^{a}\left(\boldsymbol{x}\right)\boldsymbol{u}_{\theta}^{a}, \qquad &\alpha^h_{\theta}({\bm x})=\sum_{a=1}^n N^{a}\left(\boldsymbol{x}\right)\alpha_{\theta}^{a},\nonumber \\
	&\nabla{\bm u}_{\theta}^h({\bm x})=\sum_{a=1}^n \boldsymbol{u}_{\theta}^{a}\otimes\nabla N^{a}\left(\boldsymbol{x}\right), \qquad &\nabla\alpha^h_{\theta}({\bm x})=\sum_{a=1}^n \alpha_{\theta}^{a}\,\nabla N^{a}\left(\boldsymbol{x}\right),
\end{align}
where the superscript $h$ denotes the finite element approximation of the fields, ${N}^a({\bm x})$ is the shape function corresponding to node $a$ of the finite element discretization, and $\boldsymbol{u}_{\theta}^{a}$ and $\alpha_{\theta}^{a}$ are the values of $\boldsymbol{u}_{\theta}$ and $\alpha_{\theta}$ at the same node, respectively, with $a=1...n$ and $n$ as the number of nodes.
This approach to gradient computation makes the training considerably faster.

We use the finite element discretization also for the purpose of approximating the integral in \eqref{eq6d} in order to compute the energy using the NN-based representation of the displacement and phase fields.
We evaluate the integral  employing Gauss quadrature at each element of the finite element discretization.

\subsubsection{Weight regularization}
Weight regularization plays an important role in obtaining the correct solution of a phase-field fracture problem using the same level of domain discretization as needed in FEA. In the problem of crack nucleation in a 1D bar in~\sref{bar}, when employing autodifferentiation for strain computation, the NN learns incorrect solutions with approximately zero energy after crack nucleation. An analysis of the governing PDEs (see~\ref{disp_grad_1D}) reveals that the phase-field model does not constrain the maximum slope of the displacement field near the localized phase field, thus this slope depends on the expressivity of the NN. Therefore, if the training points are not fine enough to resolve the sharply changing displacement field expressible by the NN, and not only the length scale $l$, the DRM yields incorrect solutions as shown in~\fref{figA1}. Correct results can be obtained by a finer discretization of the domain than needed to resolve the regularization length scale; this is however computationally inefficient, as it requires an even finer resolution than needed for FEA computations.  Note that, while we first recognized the issue when adopting autodifferentiation, the employment of gradient computation by finite element discretization (as illustrated in Section \ref{gradquad}) does not solve this issue. Instead, a solution is to limit $\left(\frac{{\rm d}u_{\theta}}{{\rm d}x}\right)_{max}$ by introducing weight regularization. This leads to a computationally efficient approach, in which the discretization of the domain needed for an accurate solution is comparable to the one used in FEA. Therefore, we employ weight regularization throughout the examples in this paper, see Section \ref{numSetup} for more details.

\subsubsection{Choice of the optimization algorithm}
Optimization algorithms play a key role in the training of an NN and selecting one that respects energy barriers and reaches the correct energy minimum is essential. In this study, we consider the L-BFGS algorithm~\cite{liu1989limited} and the resilient backpropagation (RPROP) algorithm~\cite{riedmiller1993direct}. While L-BFGS is the fastest of the two, it is unable to reach a good energy minimum with the localized phase field and the sharply varying displacement field (see~\ref{LBFGS_limit}). In contrast, the RPROP algorithm is observed to respect energy barriers and reach the desired energy minimum. It should be noted that the performance of first-order optimization algorithms like Adam for these problems was considerably poorer compared to L-BFGS and RPROP, hence we do not consider them further in this work.

\subsubsection{Transfer learning}
As loading is applied in incremental steps, we utilize transfer learning to learn the solution quickly. The trained network from the previous loading step is used as the initial network for training in the current loading step. We observe that transfer learning enables learning the solution field in typically fewer than 500 steps of the RPROP optimizer provided a load step does not lead to a sudden crack propagation and associated jump in energy.

\subsubsection{Nucleation and damage evolution}
Our experience shows that crack nucleation is more challenging to learn than crack propagation. In the AT1 model, the phase field does not evolve slowly with increasing load but jumps abruptly from 0 to 1 at the location of crack nucleation when the threshold loading for crack nucleation is reached. This poses a formidable challenge for NNs to learn crack nucleation in 2D. Therefore, for 2D problems involving crack nucleation we employ the AT2 model, which allows for a slow evolution of the phase field to nonzero values before it jumps to 1 at the threshold loading condition.

\subsubsection{Summary}
To summarize, our approach involves one NN with adaptive ReLU activation expressing both the displacement and phase fields. Dirichlet boundary conditions are enforced employing an ansatz which also ensures that the NN outputs corresponding to the displacement field are approximately $\sim 1$. To incentivize the phase field to lie between 0 and 1 without hindering crack nucleation, we design a new function $f_{\alpha}(\hat{\alpha}_{\theta})$. Field gradients are computed numerically like in FEA. To estimate the integral in the computation of the energy, Gauss quadrature is employed for each element in a discretized domain. In the training of the network, weight regularization is employed to constrain the maximum strain expressible by the NN. The RPROP algorithm is found to perform best in training the NN to learn the solution fields. We also employ transfer learning to learn the solution efficiently. Additionally, while the NN learns crack propagation, branching, and coalescence using the AT1 model, it is able to learn crack nucleation in 2D only with the AT2 model.

\section{Numerical examples}
\label{numE}
In this section, we first demonstrate crack nucleation in a 1D homogeneous bar and an L-shaped panel. We then study crack propagation in a single-edge notched (SEN) specimen under both tensile and shear loading conditions. To illustrate crack branching, we again model a SEN specimen under shear loading without strain energy decomposition. Crack coalescence is demonstrated by subjecting a specimen with three preexisting cracks to tensile loading.

Since these examples are taken from the phase-field fracture literature, for each example we use the same material properties as in the corresponding paper. This not only enables an easier comparison with the results of the referenced papers; it also demonstrates that the developed approach works robustly for different material properties (and geometries). 
Throughout the 2D examples, we assume plane-strain conditions.

\subsection{Numerical setup}
\label{numSetup}
For the 1D problem, we use an NN of 4 hidden layers and 50 neurons in each layer. The coefficient in the activation function is set to $m_0=1$, and we do not include $m_k$ among the learnable parameters. The L-BFGS optimizer is found to be sufficient to learn the solution in this case. Eight NNs with different initializations obtained by using Xavier initialization~\cite{glorot2010understanding} with random seeds are trained for this problem.

For the 2D problems, we use an NN of 8 hidden layers and 400 neurons in each layer. The size of the network is chosen based on the numerical experiment conducted for the problem of crack nucleation in an L-shaped panel (see~\ref{NN_sizing}). We set $m_0=2$ for crack initiation in the L-shaped panel, and $m_0=3$ for the remaining 2D problems. We train 8 NNs with different initializations for each problem employing Xavier initialization with random seeds. In the first loading step, an NN is first trained with L-BFGS before training it with RPROP; for all the subsequent loading steps only RPROP is used. For RPROP, the learning rate is set to $10^{-5}$ with the smallest and largest step sizes of $10^{-10}$ and 50, respectively. The remaining optimizer parameters are kept as in the standard implementation in Pytorch. The stopping criterion is set to correspond to a relative change in the loss function of less than $5\times10^{-6}$ for 10 consecutive optimization steps, with a maximum of 10000 optimization steps. Additionally, we apply $L_2$ weight regularization with a weight of $10^{-5}$. The Pytorch library~\cite{NEURIPS2019_9015} is used for implementation and training.

In the 1D problem, we discretize the domain into elements of size $l/5$ and assume linear shape functions. In the 2D problems, we utilize triangular elements with linear shape functions. Only the regions of the domain where the crack is expected to propagate are finely discretized with elements of size $l/5$, to limit the computational cost. Away from these regions, the element size smoothly increases up to $4l$. Discretization is performed using Gmsh~\cite{geuzaine2009gmsh}. We use one Gauss point per element for integration.

The numerical setup for the reference FEA computations is detailed in~\ref{numSetup_FEA}.

\subsection{Crack nucleation in a 1D homogeneous bar}
\label{bar}
\begin{figure}[!h]
	\begin{center}
		\includegraphics[width=0.35\textwidth]{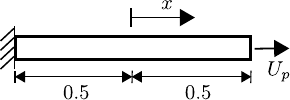}
	\end{center}
	\caption{1D bar problem: geometry and boundary conditions.}
	\label{1Dbar_dim}
\end{figure}
We first study crack nucleation in a linear elastic bar shown in \fref{1Dbar_dim}. This problem lends itself to an analysis that sheds light on the reason for the occurrence of incorrect solutions when the domain is not discretized sufficiently finely.

In this case, the expression of the energy in \eqref{eq6d} simplifies to the following form:
\begin{align}
	\mathcal{E}(u,\alpha)=\int_{-0.5}^{0.5}\frac{1}{2}g(\alpha)\left(\frac{{\rm d}u}{{\rm d}x}\right)^2{\rm d}x + \frac{1}{c_w}\int_{-0.5}^{0.5}\left(w(\alpha) +l^2\left|\frac{{\rm d}\alpha}{{\rm d}x}\right|^2\right){\rm d}x.
	\label{eq10}
\end{align}
The only parameter in the above expression is $l$, and we set $l=0.05$. We employ the AT1 model (see \eqref{eq2b}). The following ansatz is employed to apply the boundary conditions:
\begin{align}
	&u_{\theta}=[(x+0.5)(x-0.5)\hat{u}_{\theta}+(x+0.5)]U_p, \nonumber \\
	&\alpha_{\theta}=(x+0.5)(x-0.5)\hat{\alpha}_{\theta}. \label{10b}
\end{align}
Note that, since fracture at the boundaries requires less energy compared to fracture inside the domain, the NN (as well as the FEA) has a tendency to choose the solution with fracture at the boundary. To suppress this solution, we set $\alpha_{\theta}=0$ at the boundaries using the above ansatz.
\begin{figure}[!h]
	\begin{center}
		\includegraphics[width=0.49\textwidth]{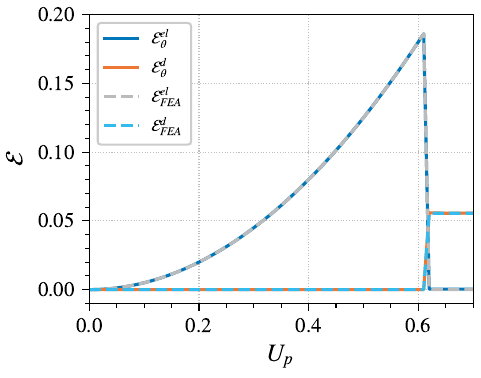}
	\end{center}
	\caption{1D bar: elastic and fracture energies vs prescribed displacement from NN and FEA.}
	\label{1Dbar_E}
\end{figure}

\begin{figure}[!h]
	\begin{center}
		\includegraphics[width=0.49\textwidth]{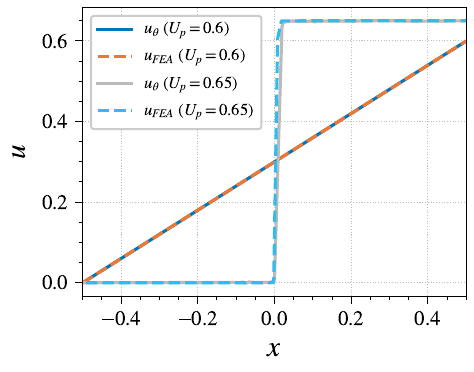}
		\includegraphics[width=0.49\textwidth]{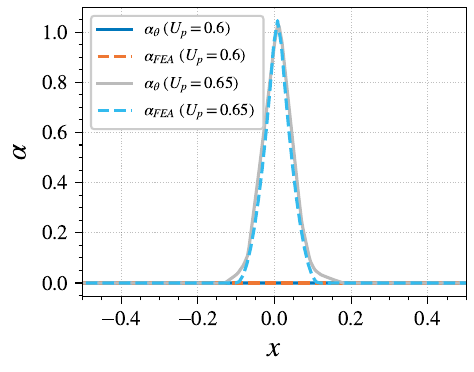}
	\end{center}
	\caption{1D bar: displacement and phase fields before and after crack nucleation from NN and FEA.}
	\label{1Dbar_field}
\end{figure}

For the 1D bar, the energies obtained from the NN are compared with the FEA solution in~\fref{1Dbar_E}, showing close agreement. Additionally, the displacement and phase fields before and after crack nucleation are compared with the FEA solution in \fref{1Dbar_field}, again showing that the DRM achieves a quite accurate solution.

\subsection{Crack nucleation in an L-shaped panel}
\label{Lpanel}
\begin{figure}[!h]
	\begin{center}
		\includegraphics[width=0.35\textwidth]{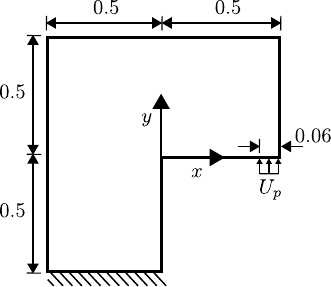}
	\end{center}
	\caption{L-shaped panel: geometry and boundary conditions.}
	\label{Lpanel_dim}
\end{figure}
This problem illustrates crack nucleation in 2D. The geometry of the panel and the boundary conditions are as shown in \fref{Lpanel_dim}, whereas the material properties are set to $\nu=0.18$ and $l=0.01$~\cite{gerasimov2019penalization}. To prescribe homogeneous and nonhomogeneous Dirichlet boundary conditions, we construct distance functions $d_1(x,y)$ and $d_2(x,y)$, respectively, such that $d_1(x,y)$ equals 1 at $y=-0.5$ and smoothly decreases to 0 away from it, and $d_2(x,y)$ equals 1 at the section of the boundary where $U_p$ is applied and smoothly decreases to 0 away from it (see~\ref{dist_funct} for more details):
\begin{align}
	& u_{\theta} = [(1-d_1(x,y))\hat{u}_{\theta}]U_p, \nonumber \\
	& v_{\theta} = [(1-d_1(x,y))(1-d_2(x,y))\hat{v}_{\theta}+d_2(x,y)]U_p, \nonumber \\ 
	& \alpha_{\theta} = f_{\alpha}(\hat{\alpha}_{\theta}). \label{eq11}
\end{align}

\begin{figure}[!h]
	\begin{center}
		\includegraphics[width=\textwidth]{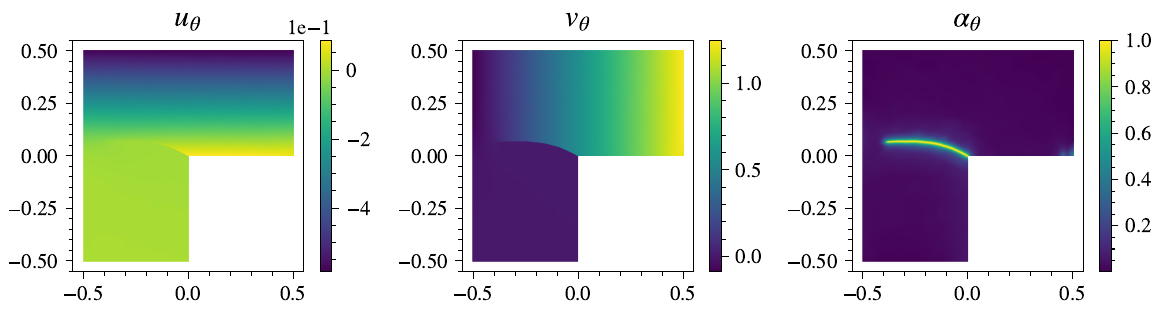}
		\includegraphics[width=\textwidth]{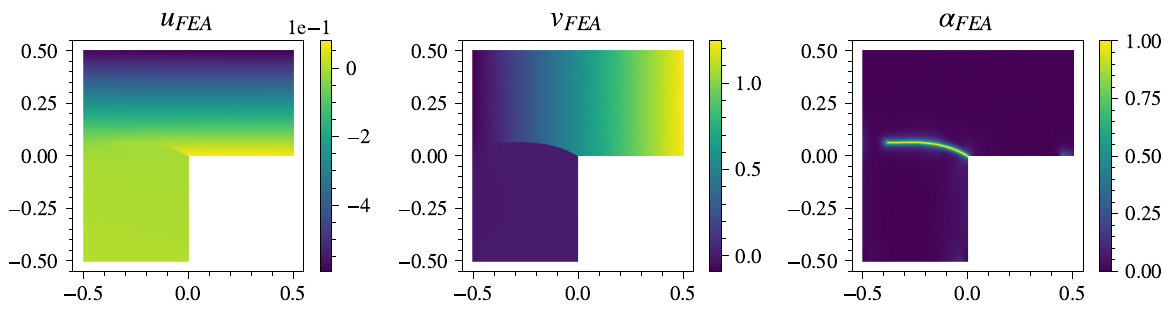}
	\end{center}
	\caption{L-shaped panel: displacement and phase fields at $U_p=1.2$ from NN and FEA.}
	\label{Lpanel_field}
\end{figure}

\begin{figure}[!h]
	\begin{center}
		\includegraphics[width=0.49\textwidth]{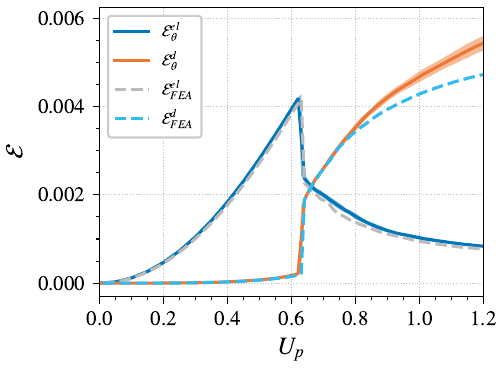}
	\end{center}
	\caption{L-shaped panel: elastic and fracture energies vs prescribed displacement from NN and FEA.}
	\label{Lpanel_E}
\end{figure}
As discussed in \sref{NNmodel}, the NN finds it difficult to learn crack nucleation when the AT1 model is employed; therefore, we use the AT2 model in this problem. \fref{Lpanel_field} compares the solution obtained from the NN with the FEA solution at $U_p=1.2$. The crack path in the NN solution closely resembles that in the FEA solution, although the FEA solution shows comparatively sharper turning of the crack. Note that, due to the non-convexity of the governing energy functional, even the FEA solution is in general not unique, and changes in the obtained crack patterns may be induced by numerical perturbations as discussed in \cite{nonunique2020}. \fref{Lpanel_E} illustrates the mean and standard deviation of the energies obtained from 8 trained NNs with different initializations, and compares them with the energies obtained from FEA. The energies are in close agreement before crack nucleation, and the NN accurately captures the critical $U_p$ at which nucleation takes place. However, in the subsequent stage the damage energy obtained from the NN is higher than in FEA. This discrepancy results from the difficulty in learning a curving crack and can be improved by reducing the loading step sizes and imposing a stricter stopping criterion for the optimizer, which however would increase the computational cost.

\subsection{Crack evolution in a notched specimen}
\label{plate_evolution}

\begin{figure}[!h]
	\begin{center}
		\includegraphics[width=0.35\textwidth]{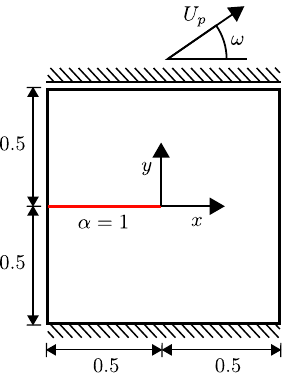}
		\hspace{4mm}
		\includegraphics[width=0.51\textwidth]{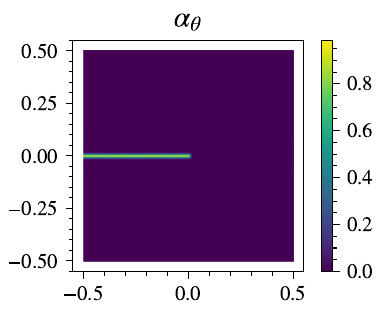}
	\end{center}
	\caption{SEN specimen: geometry and boundary conditions (left); phase field from NN at the smallest applied tensile load (right).}
	\label{SENP_dim}
\end{figure}
The geometry of the SEN specimen and boundary conditions are as shown in \fref{SENP_dim}, while the material properties are  $\nu=0.3$ and $l=0.01$~\cite{gerasimov2019penalization}. We define the following ansatz to apply the boundary conditions and to compute $\alpha$:
\begin{align}
	& u_{\theta} = [(y+0.5)(0.5-y)\hat{u}_{\theta} + (y+0.5)\cos(\omega)]U_p, \nonumber \\
	& v_{\theta} = [(y+0.5)(0.5-y)\hat{v}_{\theta} + (y+0.5)\sin(\omega)]U_p, \nonumber \\ 
	& \alpha_{\theta} = f_{\alpha}(\hat{\alpha}_{\theta}). \label{eq12}
\end{align}

We define the notch in the sample by setting an initial phase field as described in \ref{alpha_init}; the NN learns the notch as a localized phase field as shown in \fref{SENP_dim}.

\subsubsection{Tensile loading}
\label{SENP_tension}
\begin{figure}[!h]
	\begin{center}
		\includegraphics[width=\textwidth]{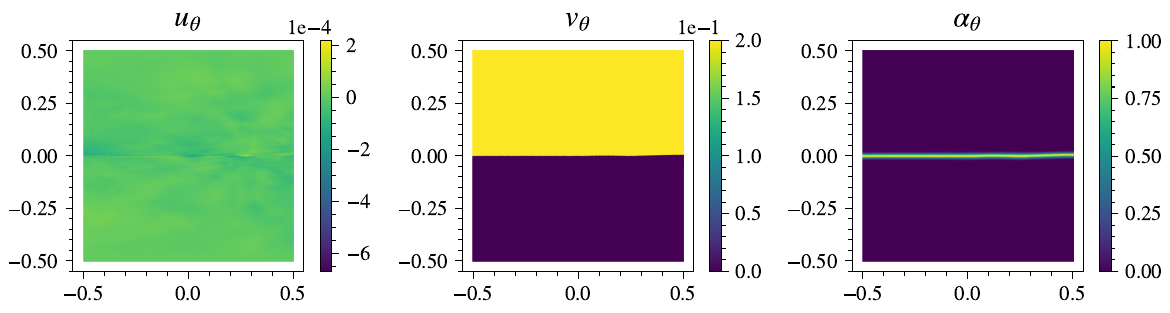}
		\includegraphics[width=\textwidth]{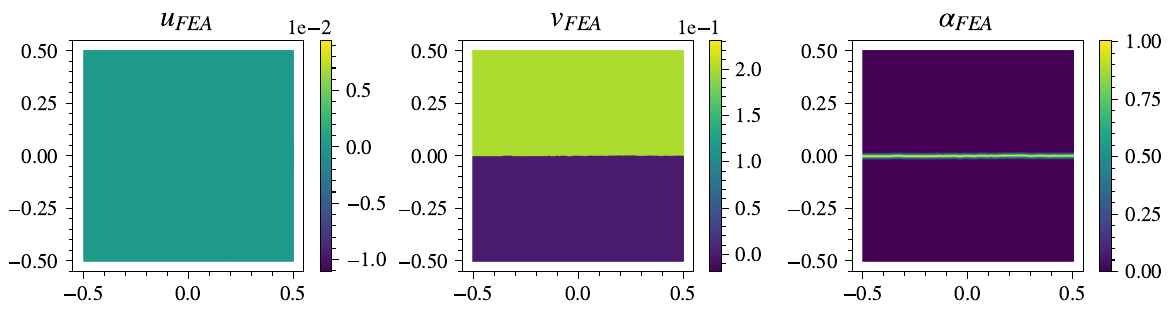}
	\end{center}
	\caption{SEN specimen under tensile loading ($\omega=\pi/2$): displacement and phase fields at $U_p=0.2$ from NN and FEA.}
	\label{SENP_tension_field}
\end{figure}
\begin{figure}[!h]
	\begin{center}
		\includegraphics[width=0.49\textwidth]{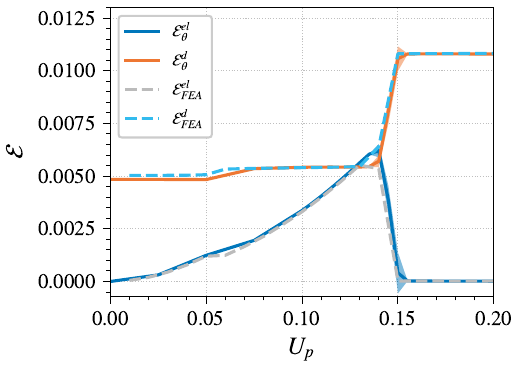}
	\end{center}
	\caption{SEN specimen under tensile loading ($\omega=\pi/2$): elastic and fracture energies vs prescribed displacement from NN and FEA.}
	\label{SENP_tension_E}
\end{figure}

With this problem, we demonstrate the ability of the NN to learn crack propagation. We subject the SEN specimen to tensile loading, i.e. the loading angle is $\omega=\pi/2$, and adopt the AT1 model. \fref{SENP_tension_field} compares the fields obtained from NN and FEA at $U_p=0.2$, showing close agreement. In this example, the simplicity of the crack path allows the NN to easily and accurately learn the cracked solution. \fref{SENP_tension_E} compares the energies associated with the NN solution with the energies from FEA, exhibiting once again a close agreement. Note that the comparatively higher standard deviation near the critical load for crack propagation results from differences in the prediction of the critical load for different initializations of the NN.

\subsubsection{Shear loading}
\label{SENP_shear}
\begin{figure}[!h]
	\begin{center}
		\includegraphics[width=\textwidth]{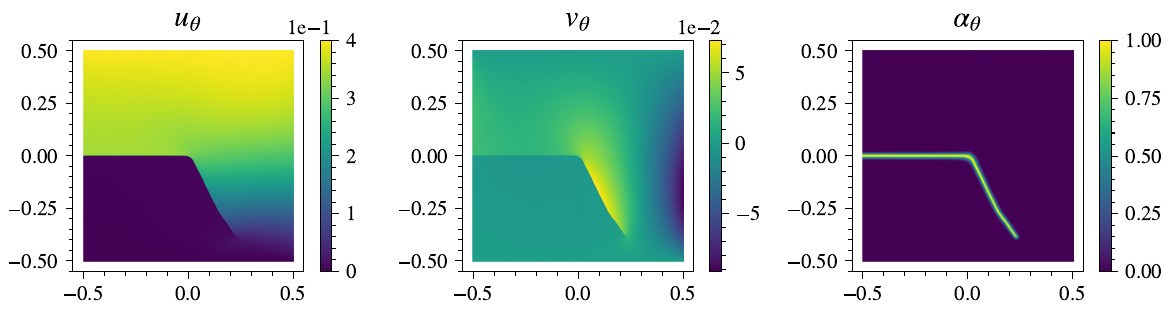}
		\includegraphics[width=\textwidth]{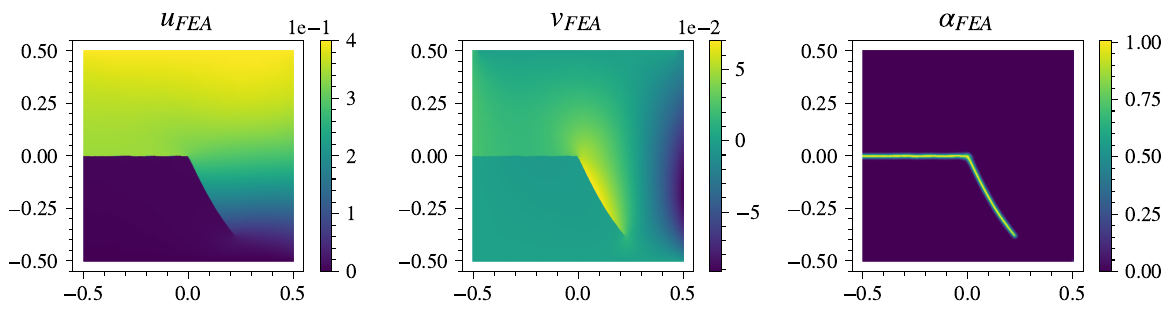}
	\end{center}
	\caption{SEN specimen under shear loading ($\omega=0$): displacement and phase fields at $U_p=0.4$ from NN and FEA.}
	\label{SENP_shear_field}
\end{figure}

\begin{figure}[!h]
	\begin{center}
		\includegraphics[width=\textwidth]{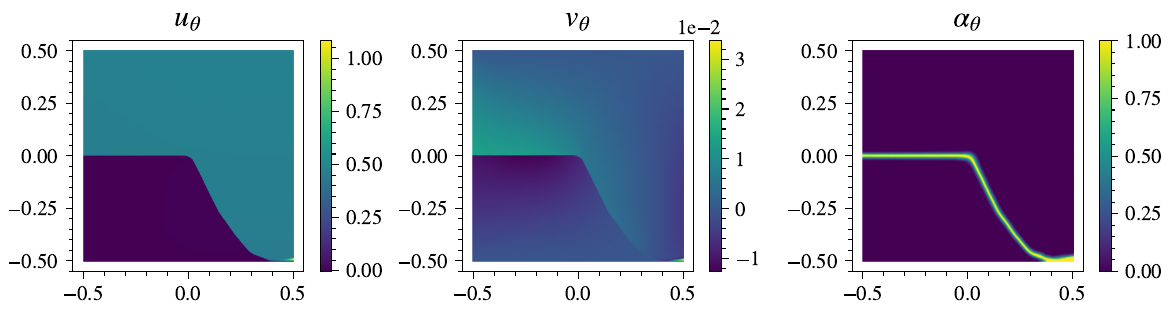}
	\end{center}
	\caption{SEN specimen under shear loading ($\omega=0$): displacement and phase fields at $U_p=0.47$ from NN.}
	\label{SENP_shear_field2}
\end{figure}

This problem demonstrates the ability of the NN to learn crack kinking, i.e. the change in direction of a crack. To this aim, we subject the SEN specimen to shear loading, i.e. the loading angle is $\omega=0$, and again use the AT1 model. The displacement and phase fields obtained from the NN at $U_p=0.4$ are compared with the FEA solution in \fref{SENP_shear_field}, showing close agreement. Note that FEA encounters convergence issues for $U_p>0.4$, while the NN does not exhibit any problem even in that range, as demonstrated by \fref{SENP_shear_field2}, which shows the fields at $U_p=0.47$.

\begin{figure}[!h]
	\begin{center}
		\includegraphics[width=0.49\textwidth]{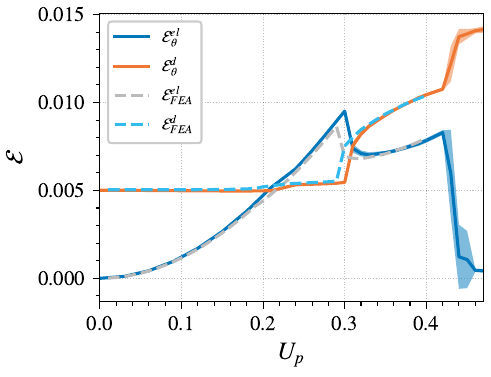}
	\end{center}
	\caption{SEN specimen under shear loading ($\omega=0$): elastic and fracture energies vs prescribed displacement from NN and FEA.}
	\label{SENP_shear_E}
\end{figure}

The energies associated with the NN solution are compared with the energies from FEA in \fref{SENP_shear_E}. They are also very close, although the critical load for crack propagation in the NN solution is slightly higher than in FEA. Moreover, higher standard deviations in the energies result from slight variations in the critical loads for different NN initializations.

\subsection{Crack branching in a notched plate}
\label{plate_branching}
\begin{figure}[!h]
	\begin{center}
		\includegraphics[width=\textwidth]{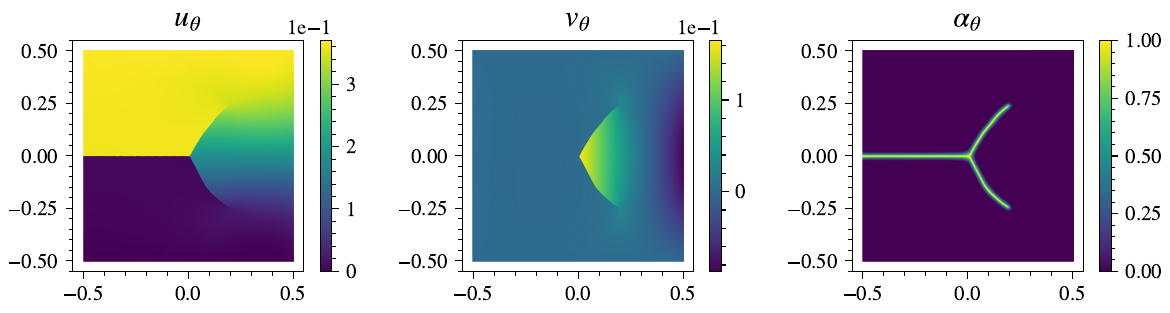}
		\includegraphics[width=\textwidth]{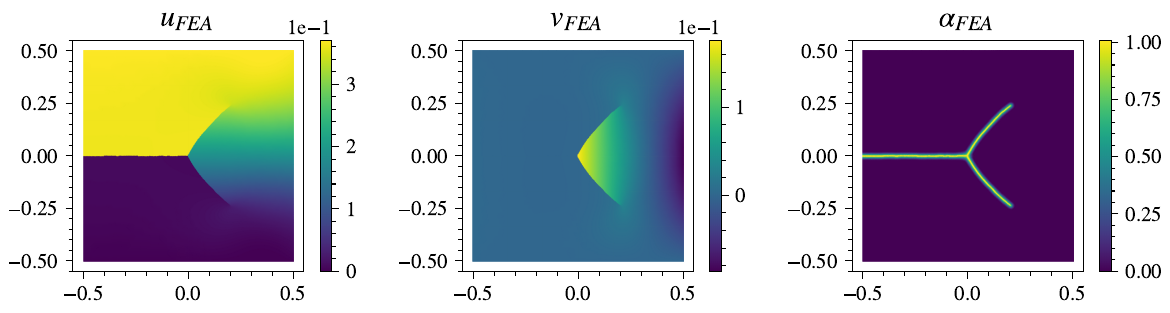}
	\end{center}
	\caption{Crack branching test: displacement and phase fields at $U_p=0.37$ from NN and FEA.}
	\label{SENP_branching_field}
\end{figure}
To demonstrate crack branching, we again prescribe the shear loading condition on the SEN specimen, i.e. $\omega=0$. In this problem, in the absence of the strain energy decomposition in \eqref{eq2d}, which differentiates between tensile and compressive stresses, unphysical branching of the crack occurs \cite{ambati2015review}. Since in this case we do not seek a physically realistic prediction of the behavior but rather an accurate solution of the governing PDEs, we purposefully omit the strain energy decomposition, i.e. we take  $\Psi^+=\Psi$ and $\Psi^-=0$. \fref{SENP_branching_field} illustrates that the NN is able to learn crack branching. The fields obtained from the NN at $U_p=0.37$ are compared with the FEA solution in~\fref{SENP_branching_field}, displaying again a good agreement, with some differences in the prediction of the curvature of the crack path.

\begin{figure}[!h]
	\begin{center}
		\includegraphics[width=0.49\textwidth]{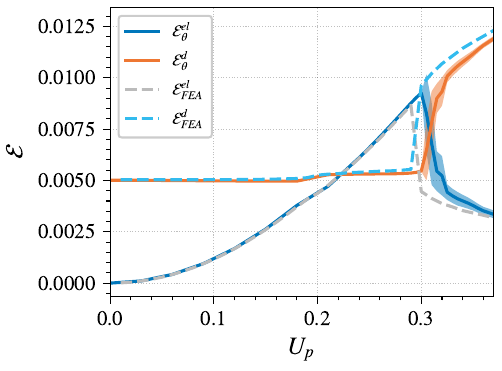}
	\end{center}
	\caption{Crack branching test: elastic and fracture energies vs prescribed displacement from NN and FEA.}
	\label{SENP_branching_E}
\end{figure}

The evolution of the energies with $U_p$ obtained from the NN are compared with the FEA energies in~\fref{SENP_branching_E}. The critical load at which branching initiates shows variability for different NN initializations, leading to a higher standard deviation; however, the average value is close to the critical load predicted by FEA. Also, the difficulty encountered by the NN in learning the curvature of the crack path is reflected in the deviation of the energies from those computed from FEA after the critical load. Note that one of the 8 NNs had difficulty in learning branching; the corresponding result is not included in the computation of the mean and standard deviation of the energies.

\subsection{Crack coalescence in a plate with cracks}
\label{plate_coalescence}
\begin{figure}[!h]
	\begin{center}
		\includegraphics[width=0.35\textwidth]{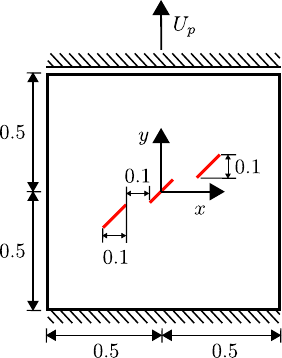}
		\hspace{4mm}
		\includegraphics[width=0.51\textwidth]{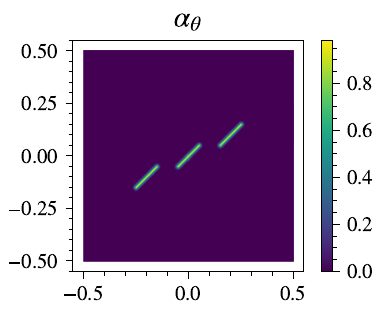}
	\end{center}
	\caption{Specimen with preexisting cracks: geometry and boundary conditions (left) and phase field at the location of the cracks from NN at the smallest applied load in training (right).}
	\label{Coalescence_dim}
\end{figure}

To demonstrate crack coalescence, tensile loading is applied on a specimen with preexisting cracks (\fref{Coalescence_dim}). In this case, we again adopt \eqref{eq12} with loading angle $\omega=\pi/2$, and the material properties are $\nu=1/3$ and $l=0.01$~\cite{zhou2018phase}. We use a larger value for $l$ in comparison to~\cite{zhou2018phase} to reduce the computational cost.

\begin{figure}[!h]
	\begin{center}
		\includegraphics[width=\textwidth]{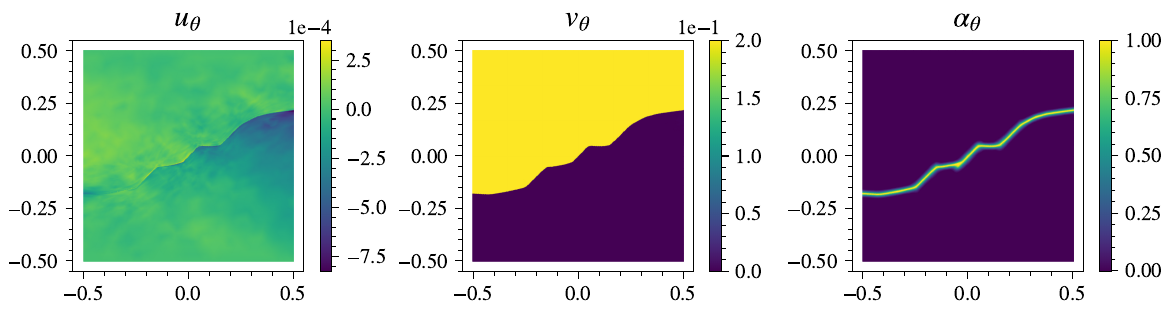}
		\includegraphics[width=\textwidth]{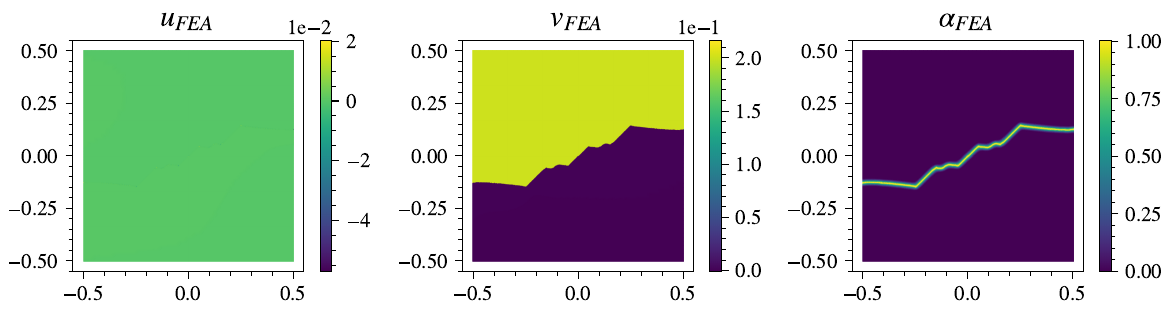}
	\end{center}
	\caption{Specimen with preexisting cracks: displacement and phase fields at $U_p=0.2$ from NN and FEA.}
	\label{Coalescence_field}
\end{figure}

\fref{Coalescence_field} demonstrates that the NN is able to learn crack coalescence. Unlike in the FEA solution, cracks starting from the initial cracks and reaching the domain boundaries are at an angle from the horizontal direction. On the other hand, as discussed earlier, the FEA solution itself is not guaranteed to be unique \cite{nonunique2020}.

\begin{figure}[!h]
	\begin{center}
		\includegraphics[width=0.49\textwidth]{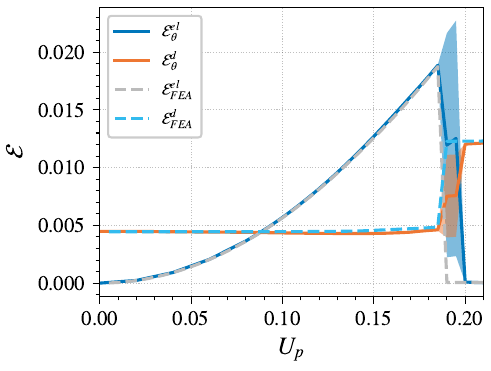}
	\end{center}
	\caption{Specimen with preexisting cracks: elastic and fracture energies vs prescribed displacement from NN and FEA.}
	\label{Coalescence_E}
\end{figure}

The energies obtained from the NN with increasing $U_p$ are compared with the FEA energies in~\fref{Coalescence_E}. The critical load at which propagation and merging occurs, while close to the FEA prediction, shows slight variability for different NN initializations, leading to a higher standard deviation near the critical load. The energies obtained from the NN after the crack propagates through the sample are in close agreement with the FEA energies even though the crack paths show differences. This suggests that different crack paths may be competing at similar energy levels, which is a known phenomenon in phase-field fracture modeling \cite{nonunique2020}. An alternative explanation may be that the NN is not reaching a sufficiently deep energy minimum, in which case results can be improved by prescribing a stricter stopping criterion for the optimizer. Note that here two NNs (out of 8) exhibit difficulty in learning the correct crack evolution, thus the corresponding results are not included in the computation of the mean and standard deviation of the energies.

\subsection{Computational cost}
\label{comp_cost}
Training of the NNs is performed on NVIDIA GeForce RTX 2080 Ti GPUs, whereas FEA computations are performed on Intel(R) Xeon(R) W-2223 CPUs.  While our DRM approach is robust and we are able to learn solutions involving various fracture phenomena, we observe that the DRM is computationally more expensive compared to FEA, with the training for NNs taking up to one order of magnitude longer than the corresponding FEA computations. This observation is not suprising; our aim in this work was not to deliver an alternative solution framework for a single boundary value problem in phase-field fracture, but  rather to develop a robust approach which can lay the foundation to solve \textit{parametric} phase-field fracture problems with the DRM.

\section{Conclusions}
\label{conclusions}
To harness physics-informed deep learning for phase-field fracture modeling, we propose the design of an NN and of a learning approach based on the DRM, aimed at learning complex fracture processes. The main conclusions of this work are as follows:
\begin{itemize}
    \item Phase-field fracture modeling requires solving a coupled problem which involves fields with sharp spatial variations. Training NNs to learn such fields is challenging; successful training relies, on one hand, on the ability of the NN to accurately approximate the energy surface and, on the other hand, on the ability of an optimizer to reach the correct local minimum respecting the energy barriers.
    
    \item We empirically establish that fully connected feed-forward NNs, with a modified ReLU activation and an appropriate ansatz in the construction of the solution fields, along with weight regularization and numerical gradient computation, is able to learn complex fracture processes. Furthermore, RPROP is found to be the most effective optimization algorithm for training.
    
    \item Our approach can solve examples of crack initiation, propagation, kinking, branching, and coalescence taken from the phase-field fracture literature, within one single numerical setup. Results in terms of solution fields and energy vs. applied displacement curves are in quantitatively excellent agreement with the respective FEA results. Additionally, our approach is robust to different network initializations.
    
\end{itemize}

In this work, we focused on the design of a robust NN and learning approach to learn the solution of boundary value problems involving diverse fracture phenomena in phase-field fracture modeling. While our approach is computationally expensive compared to FEA, our aim is not to deliver an alternative solution framework for a single boundary value problem in phase-field fracture; rather, the present work is intended as the first step in the direction of learning solutions to \textit{parametric} phase-field fracture models with the DRM. In this setting, NNs can be trained on a few realizations of the parameter space and results can then be inferred online for all other realizations, leveraging the true potential of the DRM. Learning solutions to parametric phase-field fracture models will be the goal of our future research.

\section*{Acknowledgements}
We acknowledge funding from the Swiss National Science Foundation through Grant No. 200021-219407 `Phase-field modeling of fracture and fatigue: from rigorous theory to fast predictive simulations'.

\section*{Code availability}
The code used to produce the numerical results in the paper will be made available at the time of publication.

\bibliographystyle{elsarticle-num}
\biboptions{sort&compress}
\bibliography{references}

\appendix

\section{Independent NNs for the displacement and phase fields}
\label{2NN}
\begin{figure}[!h]
	\begin{center}
			\includegraphics[width=0.4\textwidth]{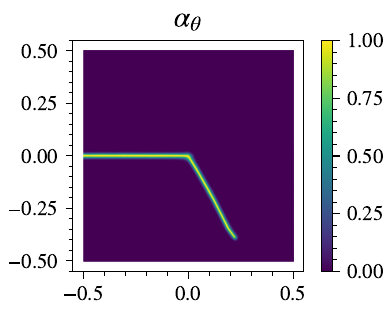}
			\hspace{2mm}
			\includegraphics[width=0.43\textwidth]{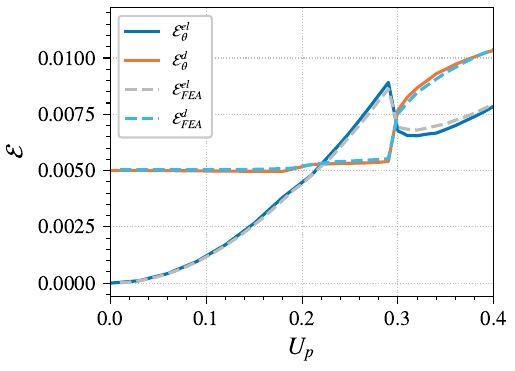}
		\end{center}
	\caption{Phase field at $U_p=0.4$ after crack propagation (left) and energies (right) for the SEN specimen under shear loading ($\omega=0$) when the displacement and phase fields are represented by two separate NNs.}
	\label{SENP_shear_2NN}
\end{figure}

\begin{figure}[!h]
	\begin{center}
		\includegraphics[width=0.4\textwidth]{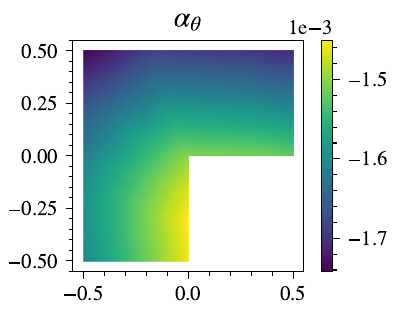}
		\hspace{2mm}
		\includegraphics[width=0.43\textwidth]{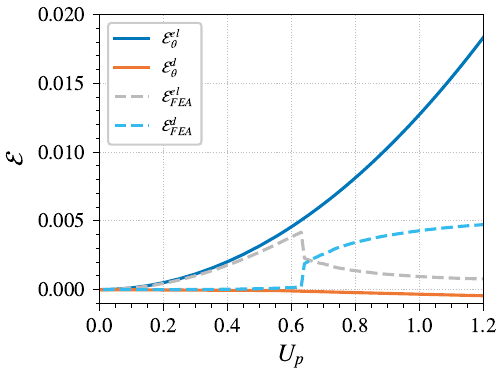}
	\end{center}
	\caption{Phase field at $U_p=1.2$ (left) and energies (right) for the L-shaped panel when the displacement and phase fields are represented by two separate NNs. In this case, the crack fails to nucleate.}
	\label{Lpanel_2NN}
\end{figure}

Using independent NNs for the physically different fields to learn their unique features separately can be expected to make training faster. We construct two NNs of 8 hidden layers and 200 neurons, representing the displacement and phase fields separately, with all the other settings as in~\sref{numE}. Note that this NN configuration lowers the number of the learnable parameters. We first study the SEN specimen under shear loading described in~\sref{SENP_shear}, and conclude that the NNs are able to learn crack propagation, as shown in~\fref{SENP_shear_2NN}. Subsequently, we use the two NNs to learn crack nucleation in an L-shaped panel as described in~\sref{Lpanel}; here, the NNs are unsuccessful, as shown in~\fref{Lpanel_2NN}. So, we use one NN to represent both the fields for all the problems.

\section{Activation functions}
\label{act}
\begin{figure}[!h]
	\begin{center}	\includegraphics[width=0.99\textwidth]{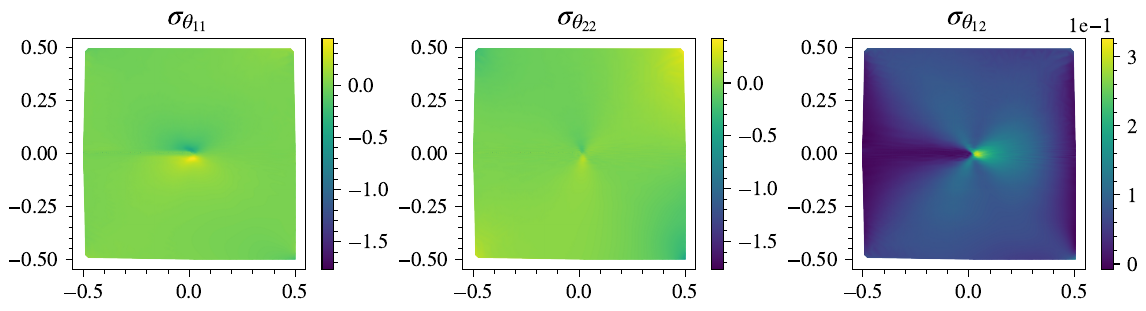}
	\end{center}
	\caption{Stress field at $U_p=0.31$ in the SEN specimen under shear loading ($\omega=0$).}
	\label{SENP_shear_stress_Tanh}
\end{figure}

\begin{figure}[!h]
	\begin{center}
		\includegraphics[width=0.4\textwidth]{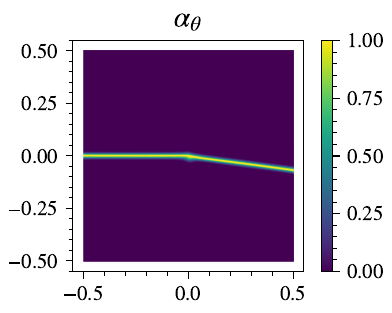}
		\hspace{2mm}
		\includegraphics[width=0.43\textwidth]{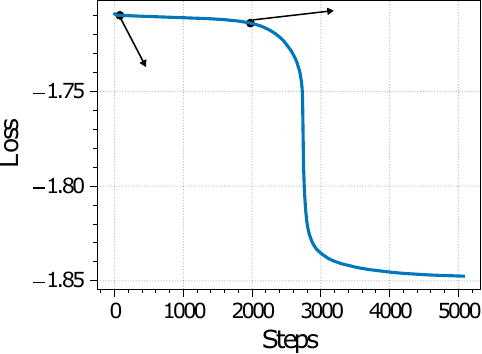}
	\end{center}
	\caption{Phase field at $U_p=0.32$ in the SEN specimen under shear loading ($\omega=0$) (left) and evolution of the loss function during training (right).}
	\label{SENP_alpha_Tanh}
\end{figure}

We assess the ability of the NNs with different activation functions in learning crack propagation in the SEN specimen under shear loading. We use the same NN size and optimizer settings as in~\sref{SENP_shear}, but set the activation function to be $\tanh$. As shown in \fref{SENP_shear_stress_Tanh}, the NN is able to learn the displacement and phase fields which yield sharp stress fields near the crack tip.
It also learns the critical $U_p$ for crack propagation accurately. However, it fails to learn the correct crack path as shown in~\fref{SENP_alpha_Tanh}. Furthermore, the smoothly decreasing loss and the comparison of the phase fields at two different optimization steps shown in~\fref{SENP_alpha_Tanh} suggest that the NN-based representation of the loss lacks the energy barrier which could prevent the NN from learning the incorrect solution.

\section{Autodifferentiation for gradient computation}
\label{diff}
\begin{figure}[!h]
	\begin{center}
		\includegraphics[width=0.43\textwidth]{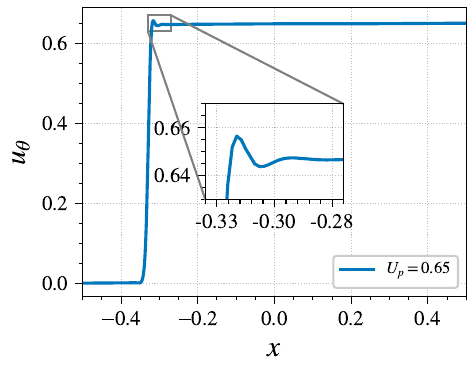}
	\end{center}
	\caption{Spurious oscillations observed in the displacement field after the localization of the phase field in the 1D bar problem.}
	\label{1D_disp_oscillation}
\end{figure}

\begin{figure}[!h]
	\begin{center}	\includegraphics[width=0.99\textwidth]{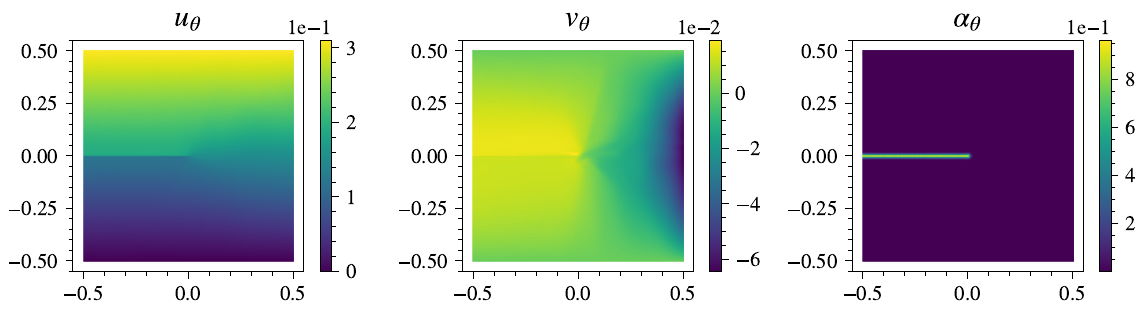}
		\includegraphics[width=0.99\textwidth]{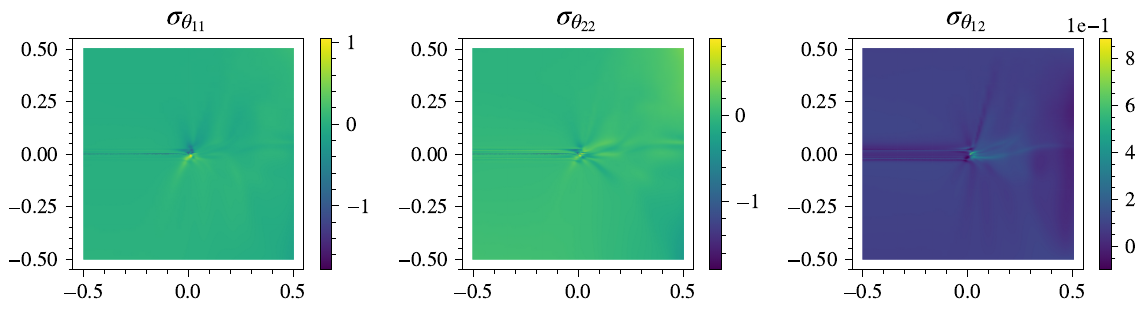}
	\end{center}
	\caption{Displacement field, phase field, and stress field at $U_p=0.31$ in the SEN specimen under shear loading ($\omega=0$).}
	\label{SENP_shear_stress_oscillation}
\end{figure}

\begin{figure}[!h]
	\begin{center}
		\includegraphics[width=0.4\textwidth]{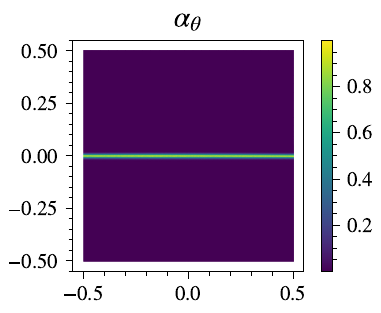}
	\end{center}
	\caption{Phase field at $U_p=0.32$ in the SEN specimen under shear loading ($\omega=0$).}
	\label{SENP_shear_alpha_incorrect}
\end{figure}

When employing a smooth activation and autodifferentiation for gradient computation, the solution field exhibits spurious oscillation. The displacement field in the 1D bar problem in~\sref{bar}, learned using an NN with 2 hidden layers and 20 neurons with the $\tanh$ activation function, exhibits oscillations as shown in~\fref{1D_disp_oscillation}.
We also train an NN with 6 hidden layers and 50 neurons with $\tanh$ activation to learn the solution of the SEN specimen under shear loading. While spurious oscillations in the displacement field are not apparent, they can be observed in the stress field as shown in~\fref{SENP_shear_stress_oscillation}. 
Furthermore, at the critical $U_p$, the crack propagates in an incorrect direction as shown in~\fref{SENP_shear_alpha_incorrect}.

\section{The maximum strain in the 1D bar}
\label{disp_grad_1D}
For the bar described in \sref{bar}, the equilibrium equation is given by
\begin{equation}
	\frac{{\rm d}\sigma}{{\rm d}x}=0, \label{eqs1}
\end{equation}
hence, the stress in the bar is uniform. Since $\sigma = g(\alpha)\frac{{\rm d}u}{{\rm d}x}$, the strain can be expressed as $\frac{{\rm d}u}{{\rm d}x} = \frac{\sigma}{g(\alpha)}=S(\alpha)\sigma$, where $S(\alpha)=1/g(\alpha)$ is the compliance. 

During evolution of the phase field, eq. \eqref{eq5c} (written as equality since $\alpha>\alpha_{n-1}$), reads
\begin{equation}
	-\frac{1}{2}S'(\alpha)\sigma^2 + w_1 \left[w'(\alpha)-2 l^2 \frac{{\rm d}^2\alpha}{{\rm d}x^2}\right] = 0, \label{eqs3}
\end{equation}
where $w_1 = \frac{1}{c_w}$ and the first term has been modified to express it in terms of compliance and stress. Multiplying the above equation with $\frac{{\rm d}\alpha}{{\rm d}x}$ and integrating with respect to $x$ yields
\begin{equation}
	-\frac{1}{2}S(\alpha)\sigma^2 + w_1 \left[w(\alpha)- l^2 \left(\frac{{\rm d}\alpha}{{\rm d}x}\right)^2\right] = C, \label{eqs4}
\end{equation}
with $C$ as the integration constant. Since, far away from the point of localization of the phase field, $\alpha$ and its gradient vanish (thus $w=0$ and $S=1$), we obtain $C = -\frac{1}{2}\sigma^2$; further substitution of $\frac{{\rm d}u}{{\rm d}x} = S(\alpha)\sigma$ yields
\begin{equation}
	\frac{{\rm d}u}{{\rm d}x}=\sigma + \frac{2w_1}{\sigma}\left[w(\alpha)-l^2\left(\frac{{\rm d}\alpha}{{\rm d}x}\right)^2 \right]. \label{eqs6}
\end{equation}

\begin{figure}[!h]
	\begin{center}
		\includegraphics[width=0.32\textwidth]{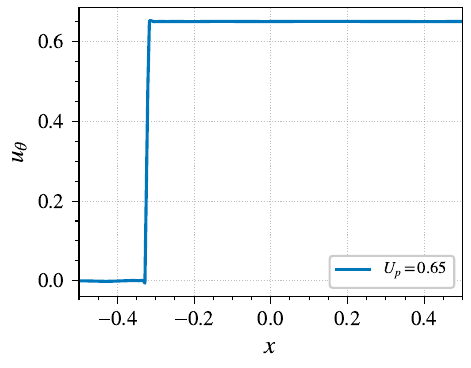}
		\includegraphics[width=0.33\textwidth]{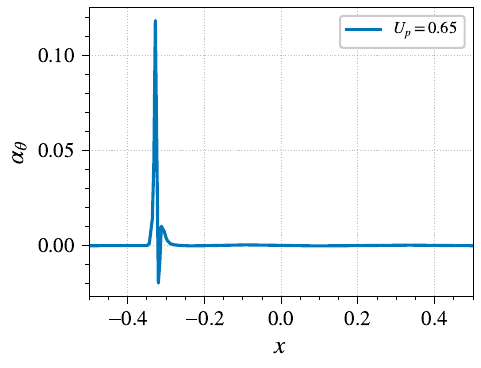}
		\includegraphics[width=0.33\textwidth]{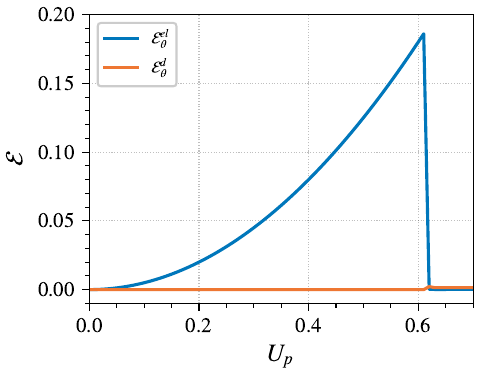}
	\end{center}
	\caption{Incorrect displacement field, phase field and energies when the maximum strain expressed by the NN solution is not controlled. The plots are obtained using an NN with 2 hidden layers and 20 neurons with the $\tanh$ activation. The correct solution fields and energies are shown in~\fref{1Dbar_field} and~\fref{1Dbar_E}, respectively.}
	\label{figA1}
\end{figure}

At the location in the bar where the maximum strain occurs, the following condition holds
\begin{equation}
	\frac{{\rm d}}{{\rm d}x}\left(\frac{{\rm d}u}{{\rm d}x}\right)=\left(w'(\alpha)-2 l^2\frac{{\rm d^2}\alpha}{{\rm d}x^2} \right)\frac{{\rm d}\alpha}{{\rm d}x}=0, \label{eqs7}
\end{equation}
hence, $\frac{{\rm d}\alpha}{{\rm d}x}=0$ and
\begin{equation}
	\left(\frac{{\rm d}u}{{\rm d}x}\right)_{max}=\sigma + \frac{2w_1 w(\alpha)}{\sigma}. \label{eqs8}
\end{equation}
When the phase field localizes, the stress in the bar goes from a positive value to 0 (however, for $\sigma=0$ the so-called optimal profile forms, where $\frac{{\rm d}\alpha}{{\rm d}x}\neq0$ at the center). When the stress approaches 0, the maximum of strain, given by \eqref{eqs8}, is unbounded. In FEA, the maximum strain is set by the element size which is chosen small enough to resolve the length scale $l$. In contrast, the maximum strain in the DRM is dictated by the expressivity of the NN. Consequently, if the training points are not fine enough to resolve the sharp slope of the displacement field expressible by an NN and not only the length scale $l$, the DRM yields an incorrect solution as shown in~\fref{figA1}.

\section{The limitations of the L-BFGS optimizer}
\label{LBFGS_limit}
\begin{figure}[!h]
	\begin{center}
		\includegraphics[width=0.99\textwidth]{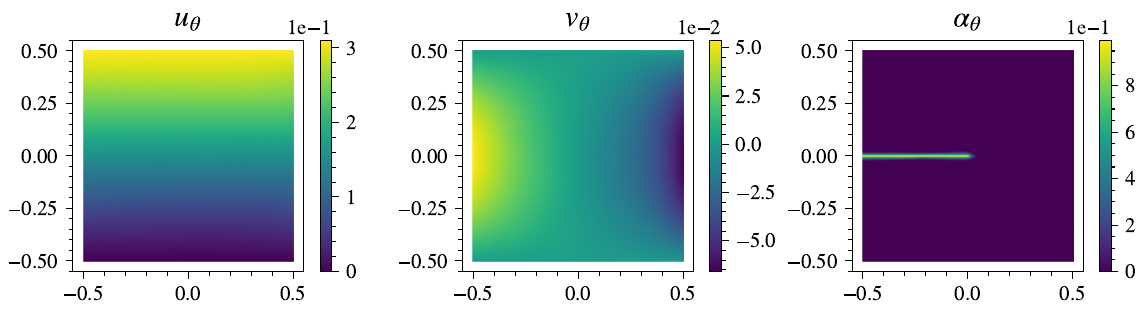}
	\end{center}
	\caption{Displacement and phase fields at $U_p=0.31$ for the SEN specimen under shear loading ($\omega=0$) when the fields are learned using only the L-BFGS optimizer.}
	\label{SENP_shear_LBFGS}
\end{figure}

To probe the ability of the L-BFGS optimizer in learning the sharp fields in the DRM, we apply it to learn the solution of the SEN sample under shear loading as described in~\sref{SENP_shear}. We train the same NN used in~\sref{SENP_shear}. As shown in~\fref{SENP_shear_LBFGS}, the NN fails to learn the sharp displacement field even before crack propagation begins. It also does not learn crack propagation as $U_p$ is increased.

\section{Network size estimation}
\label{NN_sizing}
\begin{figure}[!h]
	\begin{center}
		\includegraphics[width=0.49\textwidth]{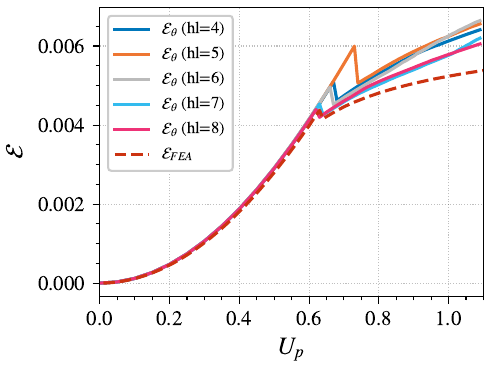}
        \includegraphics[width=0.49\textwidth]{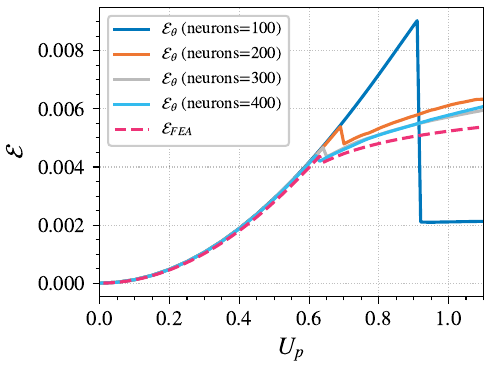}
	\end{center}
	\caption{Energy associated with the solutions learned with different network sizes vs prescribed displacement. In the left plot, number of neurons is 400; in the right plot, number of hidden layers is 8. (hl: hidden layers) }
	\label{E_NN_vary}
\end{figure}

We conduct numerical experiments with different network sizes for the problem of crack nucleation in an L-shaped panel in~\sref{Lpanel}. The energies associated with the solutions obtained from networks with different sizes are compared in~\fref{E_NN_vary}. Based on these experiments, we choose a network with 8 hidden layers and 400 neurons for all the 2D problems solved in this work.

\section{Numerical setup for FEA}
\label{numSetup_FEA}
In the FEA computation for the 1D problem in~\sref{bar},
we discretize the domain into elements of size $l/5$ and assume linear shape functions. To obtain the solution of the 2D problems, we employ quadrilateral elements with bilinear shape functions and two Gauss points per parametric direction. The regions of the domain where the crack is expected to propagate are discretized with elements of size $l/5$. The element size smoothly increases up to $4l$ away from these regions.

Initial cracks are modeled by prescribing $\alpha=1$ at the location of the crack and solving a so-called recovery problem. We set an irreversibility tolerance of $\texttt{TOL}_{ir}=10^{-3}$. The staggered solution algorithm together with the Newton-Raphson procedure is employed to iteratively converge to the solution~\cite{gerasimov2019penalization}. We set an error tolerance on the residual norm of $10^{-4}$ for the staggered scheme and $10^{-6}$ for the Newton-Raphson procedure. Moreover, we set a maximum of 500 iterations for the Newton-Raphson procedure and 1000 iterations for the staggered scheme.

\section{Distance functions for the L-shaped panel}
\label{dist_funct}
To ensure the strict enforcement of the Dirichlet boundary conditions in the problem in~\sref{Lpanel}, we construct distance functions that satisfy $C^1$ continuity. For applying the fixed boundary condition at $y=-0.5$, we construct the following distance function:
\begin{equation}
	d_1(x, y) = \begin{cases}
		\left(\frac{y+0.5}{a}-1\right)^2, & {\rm for~} y+0.5\le a \\
		0, & {\rm otherwise}
	\end{cases}
\end{equation}

To apply the prescribed displacement on a section of the boundary shown in \fref{Lpanel_dim}, we construct the following distance function:
\begin{equation}
	d_2(x, y) = \begin{cases}
		\left(\frac{|y|}{a}-1\right)^2, & {\rm for~} x\ge 0.44 {\rm ~and~} |y|\le a \\
		\left(\frac{\sqrt{(0.44-x)^2+y^2}}{a}-1\right)^2, & {\rm for~} x\le 0.44 {\rm ~and~} \sqrt{(0.44-x)^2+y^2}\le a \\
		0, & {\rm otherwise}
	\end{cases}
\end{equation}
In this work, we set $a=0.1$.

\section{Initial $\alpha$ in a notched sample}
\label{alpha_init}
\begin{figure}[!h]
	\begin{center}
		\includegraphics[width=0.25\textwidth]{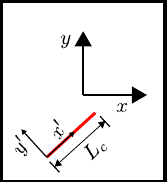}
	\end{center}
	\caption{A sample with an initial crack}
	\label{notched_sample}
\end{figure}

To define the initial $\alpha$ in a notched sample, we make use of the analytical expression for $\alpha$ in a 1D bar upon crack nucleation~\cite{wu2020phase}. Below we provide the expression for the initial $\alpha$ in a notched sample shown in \fref{notched_sample} in the AT1 model. Note that $x'-y'$ is the transformed coordinate system aligned with the crack and with its origin at one end of the crack.

\begin{equation}
	\alpha_0=\begin{cases}
		\left(1-\frac{|y'|}{2l}\right)^2, & {\rm for~} 0<x'<L_c {\rm ~and~} |y'|<2l\\
		\left(1-\frac{\sqrt{x'^2+y'^2}}{2l}\right)^2, & {\rm for~} x'<0 {\rm ~and~} \sqrt{x'^2+y'^2}<2l\\
		\left(1-\frac{\sqrt{(x'-L_c)^2+y'^2}}{2l}\right)^2, & {\rm for~} x'>L_c {\rm ~and~} \sqrt{(x'-L_c)^2+y'^2}<2l\\
		0, & {\rm otherwise}
	\end{cases}
\end{equation}

\end{document}